%% file: ms.tex
\documentclass[preprint]{emulateapj}
\slugcomment{{\sc Accepted to ApJ:} June 20, 2009}

\usepackage{natbib}
\usepackage{amsmath}

\begin{document}

\def\T1T{T_{\text{1T}}}
\def\MTT{M_{\text{2T}}}
\def\MTTe{M_{\text{2T+edge}}}
\def\MTe{M_{\text{1T+edge}}}
\newcommand{\Th}{T_{\text{hot}}}
\newcommand{\Tc}{T_{\text{cool}}}
\newcommand{\fc}{f_{\text{cool}}}
\newcommand{\normc}{\mbox{Norm}_{\text{cool}}}
\newcommand{\normh}{\mbox{Norm}_{\text{hot}}}
\newcommand{\nec}{n_{e_{\text{cool}}}}
\newcommand{\neh}{n_{e_{\text{hot}}}}
\newcommand{\rhogh}{\rho_{g_\text{hot}}}
\newcommand{\EMratio}{EM$_{\text{cool}}$/EM$_{\text{total}}$}
\newcommand{\kpc}{$h^{-1}$kpc}
\newcommand{\Mpc}{$h^{-1}$Mpc}
\newcommand{\hMsun}{$h^{-1}$M$_\odot$}
\newcommand{\Msun}{M$_\odot$}
\newcommand{\Zsun}{$Z_\odot$}
\newcommand\tna{\,\tablenotemark{a}}
\newcommand\tnb{\,\tablenotemark{b}}
\newcommand\tnc{\,\tablenotemark{c}}
\newcommand\tnd{\,\tablenotemark{d}}
\newcommand\tne{\,\tablenotemark{e}}
\newcommand\tndg{\,\tablenotemark{$\dagger$}}

\title{Discrepant Mass Estimates in the Cluster of Galaxies Abell 1689}
\shorttitle{Discrepant Mass in A1689}

\author{
E.-H.~Peng\altaffilmark{1},
K.~Andersson\altaffilmark{1},
M.~W.~Bautz\altaffilmark{1},
G.~P.~Garmire\altaffilmark{2}}

\altaffiltext{1}{MKI, Massachusetts Institute of Technology, Cambridge, MA 02139, USA}
\altaffiltext{2}{Department of Astronomy and Astrophysics, Pennsylvania State University, PA 16802, USA}
\email{epeng@mit.edu}

\begin{abstract}
We present a new mass estimate of a well-studied gravitational lensing cluster, Abell 1689, from deep {\it Chandra} observations with a total exposure of 200 ks. Within $r=200$ \kpc, the X-ray mass estimate is systematically lower than that of lensing by 30-50\%. At $r>200$ \kpc, the mass density profiles from X-ray and weak lensing methods give consistent results. The most recent weak lensing work suggest a steeper profile than what is found from the X-ray analysis, while still in agreement with the mass at large radii. Fitting the total mass profile to a Navarro-Frenk-White model, we find $M_{200}=(1.16^{+ 0.45}_{-0.27})\times10^{15}$\hMsun~with a concentration, $c_{200}=5.3^{+ 1.3}_{-1.2}$, using non-parametric mass modeling. 
With parametric profile modeling we find $M_{200}=(0.94^{+ 0.11}_{-0.06})\times10^{15}$\hMsun~and $c_{200}=6.6^{+ 0.4}_{-0.4}$.
This is much lower compared to masses deduced from the combined strong and weak lensing analysis. Previous studies have suggested that cooler small-scale structures can bias X-ray temperature measurements or that the northern part of the cluster is disturbed. We find these scenarios unlikely to resolve the central mass discrepancy since the former requires 70-90\% of the space to be occupied by these cool structures and excluding the northern substructure does not significantly affect the total mass profiles. A more plausible explanation is a projection effect. Assuming that the gas temperature and density profiles have a prolate symmetry, we can bring the X-ray mass estimate into a closer agreement with that of lensing. We also find that the previously reported high hard-band to broad-band temperature ratio in A1689, and many other clusters observed with {\it Chandra}, may be resulting from the instrumental absorption that decreases 10-15\% of the effective area at $\sim$ 1.75 keV. Caution must be taken when analyzing multiple spectral components under this calibration uncertainty. 

\end{abstract}

\section{INTRODUCTION}
Abell 1689 is a massive galaxy cluster with the largest known Einstein radius to date, $\theta_E = 45$\arcsec~for $z_s = 1$ \citep[e.g.,][]{tys90,mir95,bro05a,bro05b}, located at a moderately low redshift of $z=0.187$ \citep{fry07}. It has a regular X-ray morphology, indicating that the cluster is likely in hydrostatic equilibrium, but the mass derived from the X-ray measurement is often a factor of 2 or more lower than that from gravitational lensing at most radii. Using {\it XMM-Newton} observations, \citet[][A04~hereafter]{and04}\defcitealias{and04}{A04} find an asymmetric temperature distribution and a high redshift structure in A1689, providing evidence for an ongoing merger in this cluster. 

\cite{sah07} confirm the existance of substructures, using different sets of lensed images. This is also seen in other lensing work \citep[e.g.,][]{bro05a,die05,zek06,hal06,lim07}. Though these clumps are clearly identified, they only contribute $\simeq$ 7\% of the total mass within 250 \kpc~and are likely to be line-of-sight filaments rather than distinct merging groups. Furthermore, \cite{lok06} used the redshift distribution of galaxies to conclude that A1689 is probably surrounded by a few structures superposed along the line of sight that do not interact with the cluster dynamically, but would affect lensing mass estimates. 

A recent joint {\it Chandra}, HST/ACS, and Subaru/Suprime cam analysis by \citet[][L08~hereafter]{lem07}\defcitealias{lem07}{L08} suggested that the temperature of A1689 could be as high as $T=18$ keV at 100 \kpc, almost twice as large as the observed value at that radius. The derived 3D temperature profile was based on the X-ray surface brightness, the lensing shear, and the assumption of hydrostatic equilibrium. From the disagreement between the observed X-ray temperature and the deduced one, \citetalias{lem07} concluded that denser, colder, and more luminous small-scale structures could bias the X-ray temperature.  

In another study of 192 clusters of galaxies from the {\it Chandra} archive, \cite{cav08} find a very high hard-band (2/(1+$z$)-7 keV) to broad-band (0.7-7 keV) temperature ratio for A1689, $1.36^{+0.14}_{-0.12}$ compared to $1.16\pm0.10$ for the whole sample. They also find that merging clusters tend to have a higher temperature ratio, as predicted by \cite{mat01} where this high ratio is attributed to accreting cool subclusters lowering the broad-band temperature by contributing large amounts of line emission in the soft band.  The hard-band temperature, however, should be unaltered by this emission. The simulations of \cite{mat01} show an increase of temperature ratios of $\sim$ 20\% in general, which is close to the average of the sample of \cite{cav08}, 16\%.

A recent study, using the latest {\it Chandra} data \citep{rie08} claim that the cluster harbors a cool core and thus is relaxed based on a hardness-ratio map analysis. They further calculate a mass profile from the X-ray data and conclude that the X-ray and lensing measurements are in good agreement when the substructure to the NE is excluded. 

In this work, we examine the possibility of an extra spectral component in the X-ray data and derive an improved gravitational mass profile, including a recent 150 ks {\it Chandra} observation. \S \ref{sec_data} describes the details of data reduction and examines the possibility of an uncorrected absorption edge in the data.  In \S \ref{sec_2T}, we explore the physical properties of the potential cool substructures under a two-temperature (2T) model and examine if they can be used to explain the high hard-band to broad-band temperature ratio. In \S \ref{sec_proj}, assuming that the temperature profile derived by \citetalias{lem07} is real, we investigate what this implies for the required additional cool component. In \S \ref{sec_mass}, we derive the mass profile under both one and two temperature-phase assumptions, using both parametric and non-parametric methods. Finally, we discuss our results in \S \ref{sec_dis} and summarize in \S \ref{sec_con}.

Throughout this paper, we assume $H_0 = 100$ $h^{-1}$ km s$^{-1}$ Mpc$^{-1}$, $\Omega _{m} = 0.3$, and $\Omega _{\lambda} = 0.7$, which gives 1\arcsec~= 2.19 \kpc~at the cluster redshift of 0.187 \citep{fry07}. Abundances are relative to the photospheric solar abundances of \cite{and89}. All errors are 1 $\sigma$ unless otherwise stated.

\section{DATA REDUCTION}\label{sec_data}
\subsection{{\it Chandra}}
{\it Chandra} data were processed through CIAO 4.0.1 with CALDB 3.4.3. Since all of the observations had gone through Repro III in the archive, reprocessing data was not needed. Updated charge-transfer inefficiency and time-dependent gain corrections had already been applied. For data taken in VFAINT telemetry mode, additional screening to reject particle background was used. Events with bad CCD columns and bad grades were removed. Lightcurves were extracted from four I-chips with cluster core and point sources masked in the 0.3-12 keV band and filtered by {\it lc\_clean} which used 3 $\sigma$ clipping and a cut at 20\% above the mean. Finally, {\it make\_readout\_bg} were used to generated Out-of-Time event file. These events were multiplied by 1.3\% and subtracted from the images or the spectra to correct read-out artifacts.  For spectral anlysis, emission-weighted response matrices and effective area files were constructed for each spectral region by {\it mkacisrmf} and {\it mkwarf}.

\subsection{Background Subtraction and Modeling}\label{background}
Blank-field data sets\footnote{\url{http://asc.harvard.edu/cal/Acis/Cal\_prods/bkgrnd/acisbg}.} were used to estimate the background level. After reprojecting the blank-sky data sets onto the cluster's sky position, the background was scaled by the count rate ratio between the data and the blank-field background in the 9.5-12 keV band to account for the variation of particle induced background. Below 1 keV, the spatial varying galactic ISM emission \citep{mar03} could cause a mismatch between the real background and the blank-field data. By analysing the spectra in the same field but sufficiently far from the cluster, tailoring this soft component can be made using an unabsorbed $T\sim0.2$ keV, solar abundance thermal model \citep{vik05}. 

The current available blank-sky data were created from observations before 2005. As the solar cycle gradually reaches its minimum, the particle induced background increases. Therefore, newer observations need a much higher background normalization with a factor of 1.2-1.3. This leads an overestimate of the background in the soft band because other components like cosmic X-ray background (CXB) does not change as the particle induced background does. To correct the over-subtracted CXB and halo emission, an absorbed power law with photon index fixed at 1.4 \citep{del04} plus an unabsorbed thermal model was used to fit the blank-field background subtracted spectrum taken at $r>13$\arcmin. 

The background normalization factors used for each observation are listed in Table\,\ref{tbl-1}
\begin{deluxetable*}{cccrcccc}
\tablewidth{0pt}
\tabletypesize{\footnotesize}
\tablecaption{{\it Chandra} Observation Log\label{tbl-1}}
\tablehead{
\colhead{ObsID}&
\colhead{Data}&
\colhead{Obs. Date}&
\colhead{Exposure}&
\multicolumn{4}{c}{Background Normalization} \\  
\colhead{     }&
\colhead{Mode}&
\colhead{         }&
\colhead{(ks)    }&
\colhead{I0}&\colhead{I1}&\colhead{I2}&\colhead{I3}}
\startdata
\dataset [ADS/Sa.CXO\#obs/00540] {540}  & FAINT  & 2000-04-15 & 10.3 & 1.06 & 1.09 & 1.03 & 1.11\\
\dataset [ADS/Sa.CXO\#obs/01663] {1663} & FAINT  & 2001-01-07 & 10.7 & 1.00 & 0.98 & 0.99 & 1.04\\
\dataset [ADS/Sa.CXO\#obs/05004] {5004} & VFAINT & 2004-02-28 & 19.9 & 0.93 & 0.89 & 0.89 & 0.94\\
\dataset [ADS/Sa.CXO\#obs/06930] {6930} & VFAINT & 2006-03-06 & 75.9 & 1.21 & 1.18 & 1.18 & 1.28\\
\dataset [ADS/Sa.CXO\#obs/07289] {7289} & VFAINT & 2006-03-09 & 74.6 & 1.19 & 1.20 & 1.19 & 1.27\\[-7pt]
\enddata
\end{deluxetable*}

\subsection{{\it XMM-Newton}}
The data from two MOS detectors were processed with the XMMSAS 6.1.0 tool, {\it emchain}. Background flares were removed by a double-filtering method \citep{nev05} from $E>10$ keV and 1-5 keV light curves. Only events with pixel PATTERNs 0-12 were selected. Since {\it XMM} data were only used to crosscheck the result of the multi-component analysis of {\it Chandra} spectra, extracted from the central region where background modeling is relatively unimportant, we used the simpler local background, taken from 6\arcmin-8\arcmin. Spectral response files were created by {\it rmfgen} and {\it arfgen}. We did not include PN data because the measured mean redshift, $0.169\pm0.001$, was not consistent with those from {\it XMM} MOS or {\it Chandra} data. This could indicate a possible gain offset for PN detector, although \citetalias{and04} did not find any evidence for that.

\subsection{Systmatic Uncertainies}\label{system}
\begin{figure}
\includegraphics[width=8.5cm]{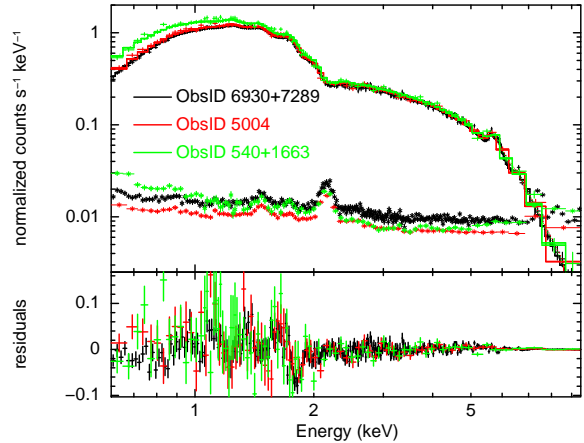}
\caption{The 0.6-9.5 keV Chandra spectrum of A1689 from the central 3\arcmin~region. The upper panel shows the data, plotted against an absorbed VAPEC model (solid line) with each element's abundance and absorption column density as free parameters. The lower panel shows residuals. 
\label{plt_spec}}
\end{figure}

\begin{figure}
\includegraphics[width=8.5cm]{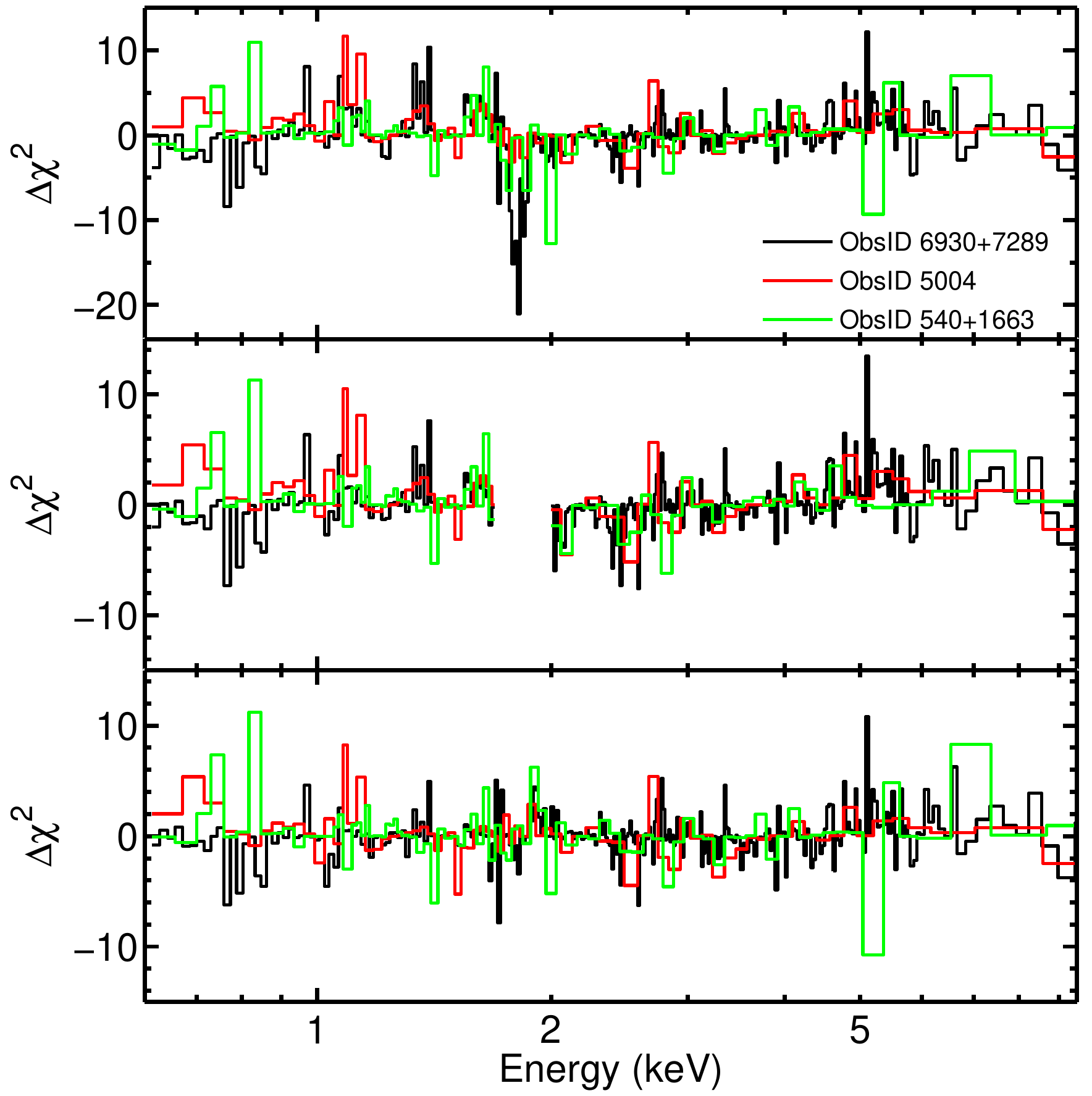}
\caption{Fit residuals, showing each channel's contribution to the total $\chi^2$. {\it Top}: an absorbed VAPEC model fit to central 3\arcmin~spectrum. {\it Middle}: same as the above, ignoring data in 1.7-2 keV. {\it Bottom}: adding an absorption edge with $E_{thresh}=$1.77 keV and $\tau$=0.12. 
\label{plt_chi}}
\end{figure}

\citetalias{lem07} pointed out some issues about previous {\it Chandra} observations (ObsID 540, 1663, and 5004). The column density from {\it Chandra} data is much lower than the Galactic value, $1.8\times10^{20}$ cm$^{-2}$ \citep{dic90}, which is also supported by the ROSAT data \citep{and04}. The temperature difference can be as high as 1.3 keV depending on the choice of column density. In the high energy band, the data is systematically higher than the model prediction. With two long {\it Chandra} observations, ObsID 6930 and 7289, we clearly see an unusual feature in the datasets which may give clues to problems mentioned before. Fig.\,\ref{plt_spec} shows an absorbed APEC model \citep{smi01} fit to the central 3\arcmin~spectrum. The prominent residual at $\sim$ 1.75 keV is present in all of our observations and appeared as the biggest contributor to the total $\chi^2$ (See Fig.\,\ref{plt_chi}). This residual can not be eliminated by adjusting individual abundances in the cluster or in the absorbing column (the cluster is at high galactic latitude). Because the residual around 1.75 keV is an order of magnitude larger than the background, it is not likely related to the background subtraction. In addition to this absorption, the residuals are systematically rising with the energy from negative to positive values. This trend is not changed when fitting the spectrum with data between 1.7-2.0 keV excluded (Fig.\,\ref{plt_chi}). We found that multiplying a XSPEC Edge model can correct the residual at $\sim$ 1.75 keV, remove steadily rising residuals with the energy, and make the column density agree with the Galactic value.

\begin{figure}
\includegraphics[width=8.5cm]{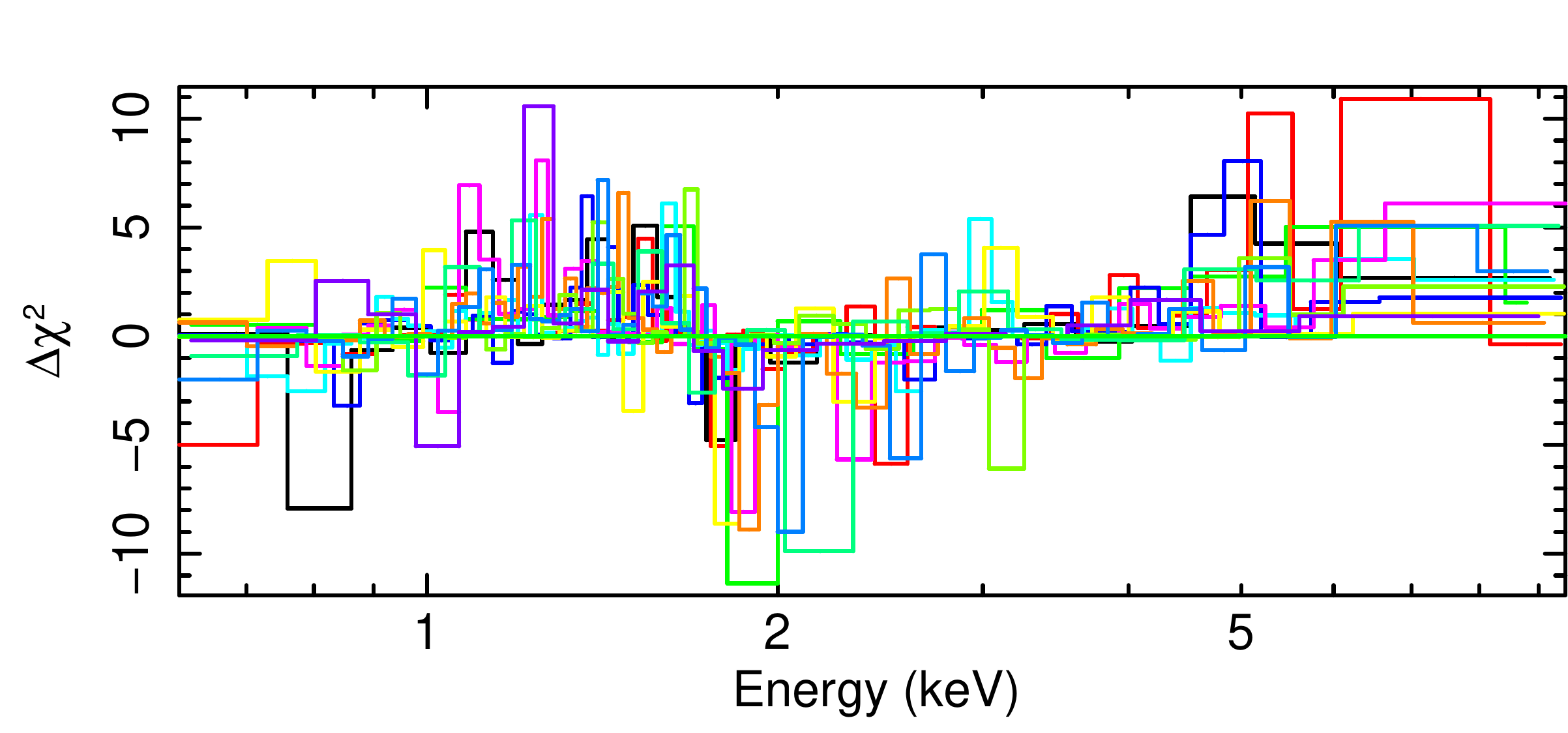}
\caption{Residuals from an absorbed APEC model fit to 12 spectra extracted from central 2.5\arcmin$\times$2.5\arcmin~ region of ObsID 6930 and 7289. We simultaneously fit these spectra and tied all parameters, except the normalization, together. 
\label{plt_boxchi}}
\end{figure}

Since the spectrum was extracted from a very large region, we averaged the position-dependent response matrices and effective area functions by the number of counts at each location. It is possible that the absorption feature is caused by improper weighting of those response files, or that this peculiarity only exists at certain regions. To dispel those doubts, we separated the central 2.5\arcmin$\times$2.5\arcmin~ area into 12 square regions and simultaneously fit these spectra with one spectral model (We only used data from ObsID 6930 and 7289 to simplify the fitting procedure). All parameters, except for the normalization, were tied together. The residuals from the single temperature fit are shown in Fig.\,\ref{plt_boxchi}. Although the fit is now acceptable with a $\chi^2/dof=3448.8/3387$, the residuals still show the same systematic trend as seen in the composite spectrum, and the column density, $(1.0^{+0.4}_{-0.4})\times10^{20}$ cm$^{-2}$ (90\% confidence level), appears low. When adding an absorption edge to the single temperature model, the derived parameters of this edge, $E_{thresh}=1.75^{+0.01}_{-0.01}$ keV and $\tau=0.15^{+0.01}_{-0.01}$, are consistent with results from the integrated spectrum. In fact, $E_{thresh}$ and $\tau$ do not strongly depend on how we model the cluster spectrum. We list fitted values of $E_{thresh}$ and $\tau$ from different cluster models and spectral extraction regions in Table\,\ref{tbl_edge}. Similar values are also found in other {\it Chandra} datasets (see the Appendix).

\begin{deluxetable}{llll}
\tablewidth{0pt}
\tabletypesize{\footnotesize}
\tablecaption{Absorption edge parameters\label{tbl_edge}}
\tablehead{
\colhead{Model}&
\colhead{fit range}&
\colhead{$E_{thresh}$}&
\colhead{$\tau$}\\
       &
       &
\colhead{(keV)}&
               }                             
\startdata
1T        & 2.5\arcmin$\times$2.5\arcmin~       & $1.75^{+0.01}_{-0.01}$ & $0.15^{+0.01}_{-0.01}$ \\
1T        & $r<$3\arcmin                        & $1.74^{+0.01}_{-0.01}$ & 
$0.14^{+0.01}_{-0.01}$ \\
1T        & $r<$3\arcmin, ignore 1.75-1.85 keV & $1.76^{+0.03}_{-0.02}$ & $0.13^{+0.01}_{-0.01}$ \\
2T\tablenotemark{a}        & $r<$3\arcmin, ignore 1.75-1.85 keV & $1.76^{+0.06}_{-0.02}$ & $0.12^{+0.02}_{-0.01}$ \\
1T+PL\tablenotemark{b}     & $r<$3\arcmin, ignore 1.75-1.85 keV & $1.76^{+0.03}_{-0.02}$ & $0.11^{+0.01}_{-0.01}$ \\
1T        & 0.2\arcmin$<r<$3\arcmin             & $1.74^{+0.01}_{-0.01}$ & $0.14^{+0.01}_{-0.01}$ \\
1T        & 0.2\arcmin$<r<$3\arcmin, ignore 1.75-1.85 keV & $1.90^{+0.04}_{-0.06}$ & $0.12^{+0.01}_{-0.01}$ \\[-7pt]
\enddata
\tablenotetext{a}{$T_1=8.0^{+0.5}_{-0.5}$ keV, $T_2=34^{+12}_{-6}$ keV.}
\tablenotetext{b}{$T_1=9.3^{+0.3}_{-0.4}$ keV, $\Gamma=-0.7^{+0.3}_{-0.4}$.}
\end{deluxetable}

The low column density can be explained by the absorption at $\sim$ 1.75 keV. This absorption has to be something like an edge, which affects the wide energy range of the spectrum, rather than an absorption line that influences only a limited energy range. The consequence to the fit resulting from this absorption is that the overall spectral normalization decreases. Since less soft photons are produced from the model, the heavy absorption by the foreground is no longer needed. We quantified the above statement in Table\,\ref{tbl_specfit} in which we simulated 100 single temperature spectra with an absorption edge at $E_{thresh}=1.74$ keV and $\tau=0.14$, fit with a single temperature (1T) model, and compared with the observations. The spectral normalization is increasing with rising column density as we exclude more data around 1.75 keV. Meanwhile, the cluster temperature and abundance are slightly decreasing. The changes of those parameters from different bandpass used in the fitting match perfectly to what are seen in the real data.

The CCD calibration around the Si-edge for ACIS-I detectors is a known issue (N. Schulz, private communication). However, it is unknown whether a correction like an edge model is needed, or if we should simply ignore the data around the Si edge. If the former is true, results from the multiple-component analysis or the hard-band/broad-band temperature measurement without applying this edge model beforehand are very questionable. As seen in Table\,\ref{tbl_specfit}, including the edge model can make the hard-band temperature 30\% hotter than the that of the broad-band. This temperature ratio depends on the cluster temperature and the quality of the data. On the other hand, if the latter is true, the spectrum implies that an additional component which is much harder than 10 keV emission is definitely required. Though, the fit is not as good as that with an edge model. From the fact that the absorption depth is sufficiently far from the zero, even if we exclude data around 1.75 keV (Table\,\ref{tbl_edge}), the intensity jump around this energy indeed exists. 

A1689 is a very hot cluster that unfortunately will be seriously affected by the calibration uncertainty around the Si edge if that can be modeled by something similar to an edge model. Lacking the knowledge that correctly treats the systematic residuals seen in the data, we provide both models, applying an absorption edge or simply ignoring the data around 1.75 keV, as our best guess to the thermal state of this cluster.

\begin{deluxetable*}{cllllcllll}
\addtolength{\tabcolsep}{-4pt}
\tabletypesize{\scriptsize}
\tablecaption{Summary of $r<3$\arcmin~spectral fits and 1T simulations \label{tbl_specfit}}
\tablehead{
\colhead{}&\multicolumn{5}{c} {Chandra observations}   & \multicolumn{4}{c} {Simulated 200 ks spectra\tna} \\
\colhead{fit range} &
\colhead{$T$} &
\colhead{$Z$} &
\colhead{$N_H$} &
\colhead{Norm} &
\colhead{$\chi^2/dof$} &
\colhead{$T$} &
\colhead{$Z$} &
\colhead{$N_H$} &
\colhead{Norm} \\
\colhead{} &
\colhead{(keV)} &
\colhead{(\Zsun)} &
\colhead{($10^{20}$cm$^{-2}$)} &
\colhead{($10^{-2}$)} &
\colhead{} &
\colhead{(keV)} &
\colhead{(\Zsun)} &
\colhead{($10^{20}$cm$^{-2}$)} &
\colhead{($10^{-2}$)} 
}
\startdata
0.6-9.5 keV\tnb
                     &$10.5^{+0.1}_{-0.1}$ & $0.36^{+0.02}_{-0.02}$ & $1.9^{+0.2}_{-0.2}$ & $1.920^{+0.007}_{-0.007}$ & 
                     1591/1390 &
                     $10.4^{+ 0.1}_{-0.1}$ & $0.36^{+0.02}_{-0.02}$ & $1.9^{+0.1}_{-0.2}$ & $1.899^{+0.008}_{-0.006}$\\
0.6-9.5 keV          &$10.7^{+0.4}_{-0.1}$ & $0.41^{+0.02}_{-0.02}$ & $0.7^{+0.2}_{-0.2}$ & $1.841^{+0.007}_{-0.007}$  &
                     1867/1390 &
                     $10.6^{+ 0.1}_{-0.1}$ & $0.40^{+0.02}_{-0.02}$ & $0.6^{+0.1}_{-0.2}$ & $1.820^{+0.008}_{-0.007}$\\
ignore 1.75-1.85 keV &$10.6^{+ 0.1}_{-0.1}$ & $0.40^{+0.02}_{-0.02}$ & $0.9^{+0.2}_{-0.2}$ & $1.857^{+0.007}_{-0.007}$ &
                     1743/1366&
                      $10.5^{+ 0.1}_{-0.1}$ & $0.39^{+0.02}_{-0.02}$ & $0.8^{+0.1}_{-0.1}$ & $1.833^{+0.008}_{-0.006}$\\
ignore 1.7-2.0 keV   &$10.5^{+0.1}_{-0.1}$ & $0.39^{+0.02}_{-0.02}$ & $1.1^{+0.2}_{-0.2}$ & $1.868^{+0.008}_{-0.008}$ &
                     1645/1327 &
                      $10.3^{+ 0.1}_{-0.1}$ & $0.38^{+0.02}_{-0.02}$ & $1.2^{+0.2}_{-0.1}$ & $1.852^{+0.008}_{-0.007}$\\
ignore 1.7-2.5 keV   &$10.4^{+0.1}_{-0.1}$ & $0.38^{+0.02}_{-0.02}$ & $1.6^{+0.2}_{-0.2}$ & $1.892^{+0.008}_{-0.008}$ &
                     1466/1222 &
                     $10.2^{+ 0.1}_{-0.1}$ & $0.37^{+0.02}_{-0.01}$ & $1.6^{+0.2}_{-0.2}$ & $1.877^{+0.007}_{-0.008}$\\
ignore 1.7-3.0 keV   &$10.4^{+0.1}_{-0.1}$ & $0.37^{+0.02}_{-0.02}$ & $1.8^{+0.2}_{-0.2}$ & $1.904^{+0.009}_{-0.009}$ &
                     1317/1120 &
                     $10.2^{+ 0.1}_{-0.1}$ & $0.36^{+0.02}_{-0.01}$ & $1.7^{+0.2}_{-0.2}$ & $1.886^{+0.008}_{-0.009}$\\ 
1.7-7.0 keV          &$13.1^{+0.3}_{-0.3}$ & $0.44^{+0.02}_{-0.02}$ & 1.8 & $1.817^{+0.008}_{-0.008}$ & 
                     1100/960 &
                     $12.7^{+ 0.3}_{-0.3}$ & $0.44^{+0.02}_{-0.02}$ & 1.8 & $1.790^{+0.008}_{-0.008}$\\
0.7-7.0 keV          &$10.1^{+0.1}_{-0.1}$ & $0.40^{+0.02}_{-0.02}$ & 1.8 & $1.867^{+0.006}_{-0.006}$ & 
                     1655/1164 &
                     $10.0^{+ 0.1}_{-0.1}$ & $0.39^{+0.02}_{-0.02}$ & 1.8 & $1.849^{+0.005}_{-0.006}$\\[-7pt]
\enddata
\tablenotetext{a}{The absorption edge is at $E_{thresh}=1.74$ keV with $\tau=0.14$. We used $T=10.5$ keV, $Z=0.36$ \Zsun, $z=0.187$, Norm $=1.9\times10^{-2}$ (corresponding to $S_{X[0.6-9.5keV]}=2.4\times10^{-11}$ erg cm$^{-2}$s$^{-1}$), $N_{H}=1.8\times10^{20}$ cm$^{-2}$. } 
\tablenotetext{b}{Multiplied by an absorption edge at $E_{thresh}=1.74$ keV with $\tau=0.14$.}
\tablecomments{Errors are 1$\sigma$ or 68\% CL for 100 simulations.}
\end{deluxetable*}

\subsection{Spectral fitting}\label{fitting}
Spectra were fit with XSPEC  12.3.1 package \citep{xspec}. We adopted $\chi^2$ statistic and grouped the spectra to have a minimum of 25 counts per bin.  However, when fitting background dominated spectra, $\chi^2$ statistic is proven to give biased temperature \citep{lec07}. Another choice available in XSPEC is using Cash statistic with modeled, rather than subtracted background. Since modeling the background needs many components: cosmic ray induced background (broken power-laws plus several Gaussians), particle background (broken power-laws), cosmic X-ray background (power-law), galactic emission (thermal), etc, the whole spectral model will be very complicated for analysis like deprojection, which simultaneously fits all of the spectra extracted at different radii. We decided to use $\chi^2$ statistic but with a different grouping method to bypass the difficulty in background modeling. 

As shown in Table\,\ref{tbl_chi}, we simulated 500 {\it Chandra} spectra with $N_H=1.8\times10^{20}$ cm$^{-2}$, $Z=0.2$ solar, and $T=9, 7, 5$ keV with Norm\footnote{Spectral normalization, $\mbox{Norm} = \frac{10^{-14}}{4\pi((1+z)D_A)^2}\int n_en_HdV$, where $n_e$ and $n_H$ are in cm$^{-3}$, $V$ in cm$^3$, and $D_A$ in cm.} $=1.51\times10^{4}, 1.52\times10^{4}, 1.56\times10^{4}$, respectively.  The spectral normalization was chosen to match the observed flux at $r=$ 6.5\arcmin-8.8\arcmin, where the background is $\sim 90$\% of the source in 0.9-7.0 keV band. Spectra were generated based on the response files of ObsID 6930 and 7289 with a total exposure time of 150 ks. When data are binned to have a minimum of 25 total counts (background included) per channel, a 9 keV gas will be measured to be 5 keV. Raising the threshold can lessen this bias. However, even with 400 counts per bin, which greatly reduces the spectral resolution by a factor of 10, the temperature is still being underestimated by $\sim$ 1 keV for $T=9$ keV gas.  We found out that binning data to have at least 2 counts above the background can recover the true temperature, though this minimum have to be adjusted according to the background contribution. Spectra at large radii were binned by this grouping scheme.

\begin{deluxetable}{lccc}
\tablewidth{0pt}
\tabletypesize{\footnotesize}
\tablecaption{Summary of 1T simulations of 150 ks {\it Chandra} spectra at $r=$ 6.5\arcmin-8.8\arcmin.  \label{tbl_chi}}
\tablehead{\colhead{Min counts}     & 
           \colhead{$T_0$ (keV)}    &  
           \colhead{$T_{med}$ (keV)}& 
           \colhead{$dof_{mean}$}   \\
           \colhead{(1)}            &
           \colhead{(2)}            &
           \colhead{(3)}            &
           \colhead{(4)}             }
\startdata
25 (tot) & 9 & $ 4.9^{+ 2.1}_{-1.2}$ &  396\\
         & 7 & $ 4.2^{+ 1.6}_{-1.1}$ &  396\\
         & 5 & $ 3.4^{+ 0.9}_{-0.7}$ &  395\\
\hline
100 (tot)& 9 & $ 6.8^{+ 5.1}_{-1.8}$ &  142\\
         & 7 & $ 5.9^{+ 3.1}_{-1.7}$ &  142\\
         & 5 & $ 4.3^{+ 1.5}_{-1.1}$ &  141\\
\hline
400 (tot)& 9 & $ 7.9^{+ 7.2}_{-2.4}$ &   40\\
         & 7 & $ 6.6^{+ 4.2}_{-2.0}$ &   40\\
         & 5 & $ 4.8^{+ 1.9}_{-1.3}$ &   40\\
\hline
2 (net)  & 9 & $ 8.6^{+ 7.8}_{-2.7}$ &  170\\
         & 7 & $ 7.2^{+ 4.7}_{-2.2}$ &  166\\
         & 5 & $ 5.2^{+ 1.9}_{-1.4}$ &  155\\[-7pt]
\enddata
\tablecomments{(1) minimum (total or net) counts per channel, (2) input temperature, (3) median temperature, (4) median $dof$. Errors are 68\% CL for 500 simulations.}
\end{deluxetable}

\section{SPECTRAL ANALYSIS}\label{sec_2T}
\subsection{Single temperature model}\label{1T}
We first determined the general properties of the cluster, using the spectrum extracted from the central 3\arcmin~(395 \kpc) region and fitting it with a single temperature VAPEC model. The Ne, Mg, Si, S, Ar, Ca, Fe, Ni abundances and the redshift were free to vary. The column density was fixed at the Galactic value. The best-fit parameters are listed in Table\,\ref{tbl_spec1t}. For {\it Chandra} data we fixed the Si abundance at 0.4 \Zsun, since the residual at 1.75 keV (\S \ref{system}) was close to  Si \textsc{xiv} K$\alpha$ line (2.01 keV, rest-frame).

Table\,\ref{tbl_spec1t} shows that a single-temperature model is moderately adequate for {\it XMM-Newton} MOS data but not for {\it Chandra} when the absorption edge is not modeled. In addition to this difference, the {\it Chandra} temperature is $\sim$1 keV higher than that of {\it XMM-Newton}. This temperature disagreement is likely related to the cross-calibration problems, as noted in other studies \citep[e.g.,][L. David\footnote{\url{http://cxc.harvard.edu/ccw/proceedings/07\_proc/presentations/david}}]{kot05,vik06,sno08}, but it could also be caused by incorrect cluster modeling. A two-temperature model will be investigated in \S\ref{2T}.

{\it Chandra} data also have a higher Fe and a much higher Ni abundance, resulting an unusually high Ni/Fe ratio of $ 7.5\pm1.3$ and $9.0\pm1.2$ Ni$_{\sun}$/Fe$_{\sun}$ with and without an absorption edge correction, respectively, in contrast to the {\it XMM-Newton} value of $1.8\pm1.6$ Ni$_{\sun}$/Fe$_{\sun}$. Our {\it XMM-Newton} MOS result is in agreement with that of \cite{dep07}, $0.9\pm1.5$ Ni$_{\sun}$/Fe$_{\sun}$, obtained from MOS and PN spectra from the $r<1.3$\arcmin~region with a differential emission measure MEKAL-based model, {\it wdem} \citep{kaa04}. Such a high Ni/Fe ratio greatly exceeds the yield of typical SN Ia models \citep{iwa99}, which range from 1.4-4.8 Ni$_{\sun}$/Fe$_{\sun}$. Since there is a known temperature discrepancy between {\it Chandra} and {\it XMM-Newton} that would affect elemental abundance determinations, direct Fe and Ni line measurements will be conducted in \S\ref{line}.

The S abundance, determined mostly by the S \textsc{xvi} K$\alpha$ line at 2.62 keV (rest-frame), should be accurately measured for {\it XMM-Newton} EPIC since it suffers little systematic uncertainty \citep{wer08}. However, without an absorption edge correction, there is basically no S detection for {\it Chandra} data, which strongly contradicts the {\it XMM-Newton} result. This shows the great impact of the absorption at 1.75 keV. When including an edge model into the fit, we have consistent S abundances for both instruments.

\begin{deluxetable*}{cccccccccccl}
\addtolength{\tabcolsep}{-10pt}
\tabletypesize{\scriptsize}
\tablecaption{Best-fit VAPEC parameters\label{tbl_spec1t}}
\tablehead{
\colhead{} & \colhead{$T$ (keV)} & \colhead{$z$} & \colhead{Ne} & \colhead{Mg} & \colhead{Si} & \colhead{S} & \colhead{Ar} & \colhead{Ca} & \colhead{Fe} & \colhead{Ni} & \colhead{$\chi^2/dof$}}

\startdata
Chandra \tablenotemark{a}& $10.2^ { +0.1 } _ { -0.1 } $ & $0.186^ { +0.001 } _ { -0.001 } $ & $0.59^ { +0.32 } _ { -0.29 } $ & $1.95^ { +0.30 } _ { -0.30 } $ & $0.4 _ { f }         $ & $<0.05                       $ & $<0.19                       $ & $<0.12                       $ & $0.40^ { +0.02 } _ { -0.02 } $ & $3.60^ { +0.44 } _ { -0.47 } $ &  1664/1360 \\ 
Chandra \tablenotemark{b} & $10.5^ { +0.1 } _ { -0.1 } $ & $0.186^ { +0.001 } _ { -0.001 } $ & $<0.18                       $ & $1.16^ { +0.22 } _ { -0.42 } $ & $0.4 _ { f }         $ & $0.66^ { +0.22 } _ { -0.32 } $ & $<0.38                       $ & $<0.09                       $ & $0.37^ { +0.01 } _ { -0.02 } $ & $2.77^ { +0.48 } _ { -0.45 } $ &  1545/1382 \\ 
XMM MOS & $9.4^ { +0.1 } _ { -0.1 } $ & $0.183^ { +0.001 } _ { -0.001 } $ & $0.35^ { +0.40 } _ { -0.35 } $ & $1.55^ { +0.43 } _ { -0.40 } $ & $0.50^ { +0.20 } _ { -0.19 } $ & $0.64^ { +0.23 } _ { -0.23 } $ & $<0.57                       $ & $1.38^ { +0.69 } _ { -0.71 } $ & $0.31^ { +0.02 } _ { -0.02 } $ & $0.57^ { +0.49 } _ { -0.47 } $ &   982/840 \\[-7pt]
\enddata
\tablecomments{Al, O fixed at 0.4 solar and He, C, N at 1 solar. For the elements whose abundances reach the lower bound, zero, only the upper limits are shown. Errors are 1$\sigma$.}
\tablenotetext{a}{without an absorption edge and ignoring data at 1.75-1.85 keV.}
\tablenotetext{b}{with an absorption edge. Edge parameters are determined from the data with  $E_{thresh}=1.74^{+0.01}_{-0.01} $ keV and $\tau=0.13^{+0.01}_{-0.01}$.} 
\end{deluxetable*}

\subsection{Two temperature model}\label{2T}
To get some clues to the nature of the claimed cool substructures in A1689, a simple two temperature model was fit to the spectrum extracted from the $r<3$\arcmin~(395 \kpc) region where the quality of the data was high enough to test it. We used two absorbed VAPEC models, with variable normalization but linked metallicities between the two phases. The column density was fixed at the Galactic value. To reduce the uncertainty on measuring metallicities, we tied the abundances of $\alpha$-elements (O, Ne, Mg, Si, S, Ar, and Ca) together and fixed the remaining abundances at the solar value, except for Fe and Ni. Since the hotter phase temperature, $\Th$, was harder to constrain, it was frozen at a certain value above the best-fit single temperature fit, $\T1T$. We changed this increment from 0.5 to 50 keV to explore the whole parameter space.

\begin{figure}
\epsscale{1.0}
\includegraphics[width=8.5cm]{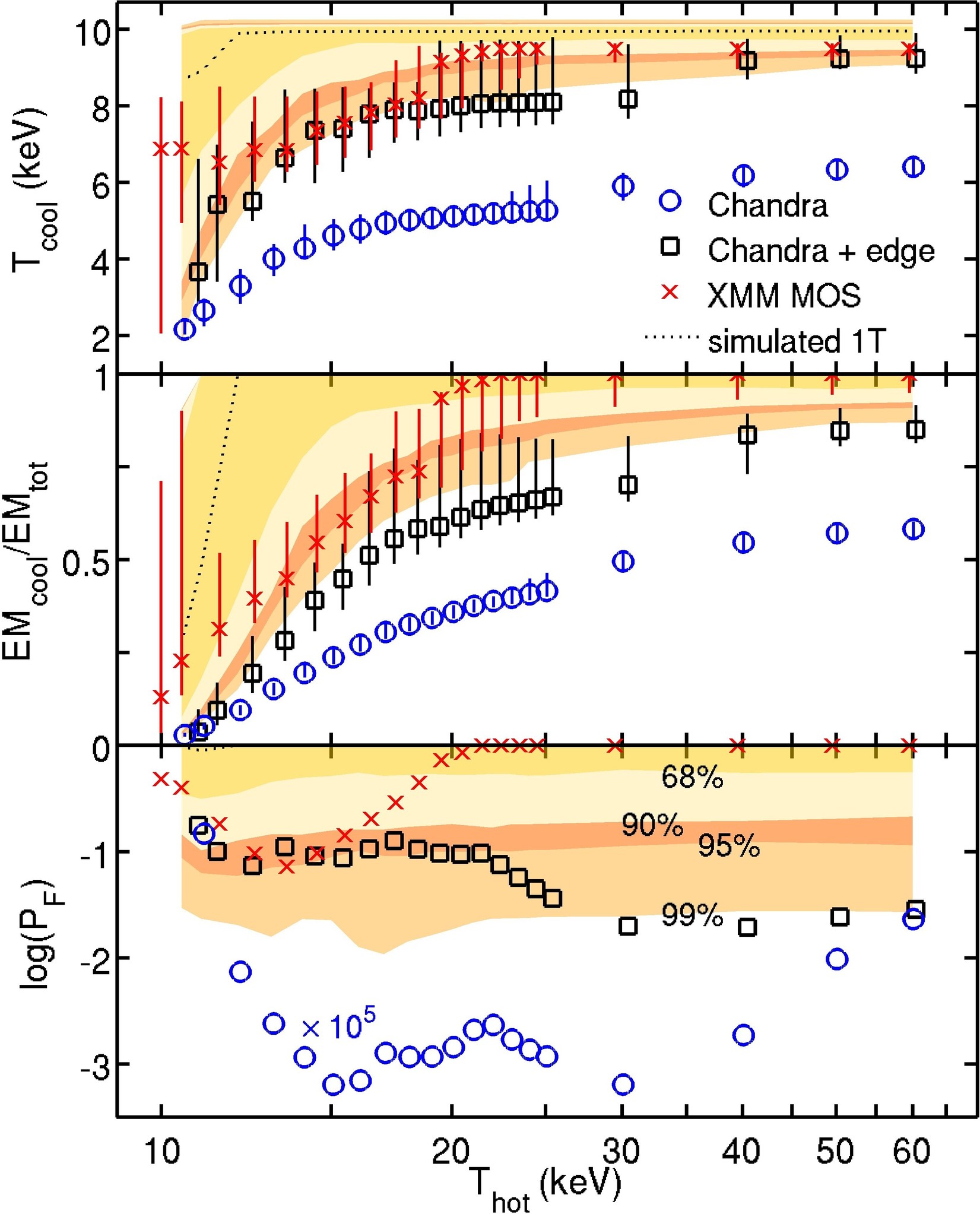}
\caption{The temperature of the cooler gas $\Tc$, the emission measure ratio \EMratio, and the F-test probability $P_F$ are plotted as a function of $\Th$. The shaded region represents 68\%, 90\%, 95\%, and 99\% CL from 1000 simulated $T=10.1$ keV {\it Chandra} spectra. The $P_F$ from Chandra data without the absorption edge corrected (circles) is multiplied by $10^5$.
\label{plt_2t_chi}}
\end{figure}

Fig.\,\ref{plt_2t_chi} shows the temperature of the cooler gas, $\Tc$, and the fractional contribution of the cooler gas, \EMratio,  as a function of $\Th$.  As $\Th$ increases, $\Tc$ and \EMratio increase as well. $\Tc$ eventually becomes $\T1T$ once $\Th$ is greater than 20 keV and very little gas is left in the hot phase, which is also supported by the {\it XMM-Newton} data.  For $\Th$ $\approx$ 18 keV, there has to be 30\%, 60\% of the cool gas at the temperature of 5, 8 keV inferred from {\it Chandra} and {\it XMM-Newton} data, respectively. {\it Chandra} absorption corrected data show similar results as {\it XMM-Newton} data do at this temperature. Although there is some inconsistency between {\it Chandra} and {\it XMM-Newton} data, both indicate that the cool component, if it indeed exists, is not cool at all. $T=5$ keV is the typical temperature of a medium sized cluster with a mass of $M_{500}=2.9\times10^{14}h^{-1}$\Msun \citep{vik06}.

To quantify how significant the detection of this extra component was, we conducted an $F$-test from the fits of 1T (the null model) and 2T models. However, because the 2T model reduces to 1T when the normalization of one of the two components hits the parameter space boundary (ie, zero), the assumption of $F$-test is not satisfied \citep[see][]{pro02}. Therefore, we simulated 1000 1T {\it Chandra} spectra and performed the same procedure to derive the $F$-test probability, $P_F$, based on the $F$ distribution. Fig.\,\ref{plt_2t_chi} shows the distribution of $P_F$ from simulated data at the 68, 90, 95, and 99 percentile overplotted with $P_F$ from {\it Chandra} and {\it XMM-Newton} data.  We plot $P_F$ in Fig.\,\ref{plt_2t_chi} rather than the $F$ statistic, since $P_F$ is a scaler that does not depend on the degrees of freedom of the fits and is ideal to compare observations that have different data bins. For $\Th < 20~$keV, both the edge-corrected {\it Chandra} data and the {\it XMM-Newton} data are within the 95 percentile of the simulated 1T model and we conclude that a 2T model is possible but not necessary to describe the data.

\subsection{Hard-band, broad-band temperature}\label{Thbr}
\begin{figure}
\begin{center}
\includegraphics[width=7.5cm]{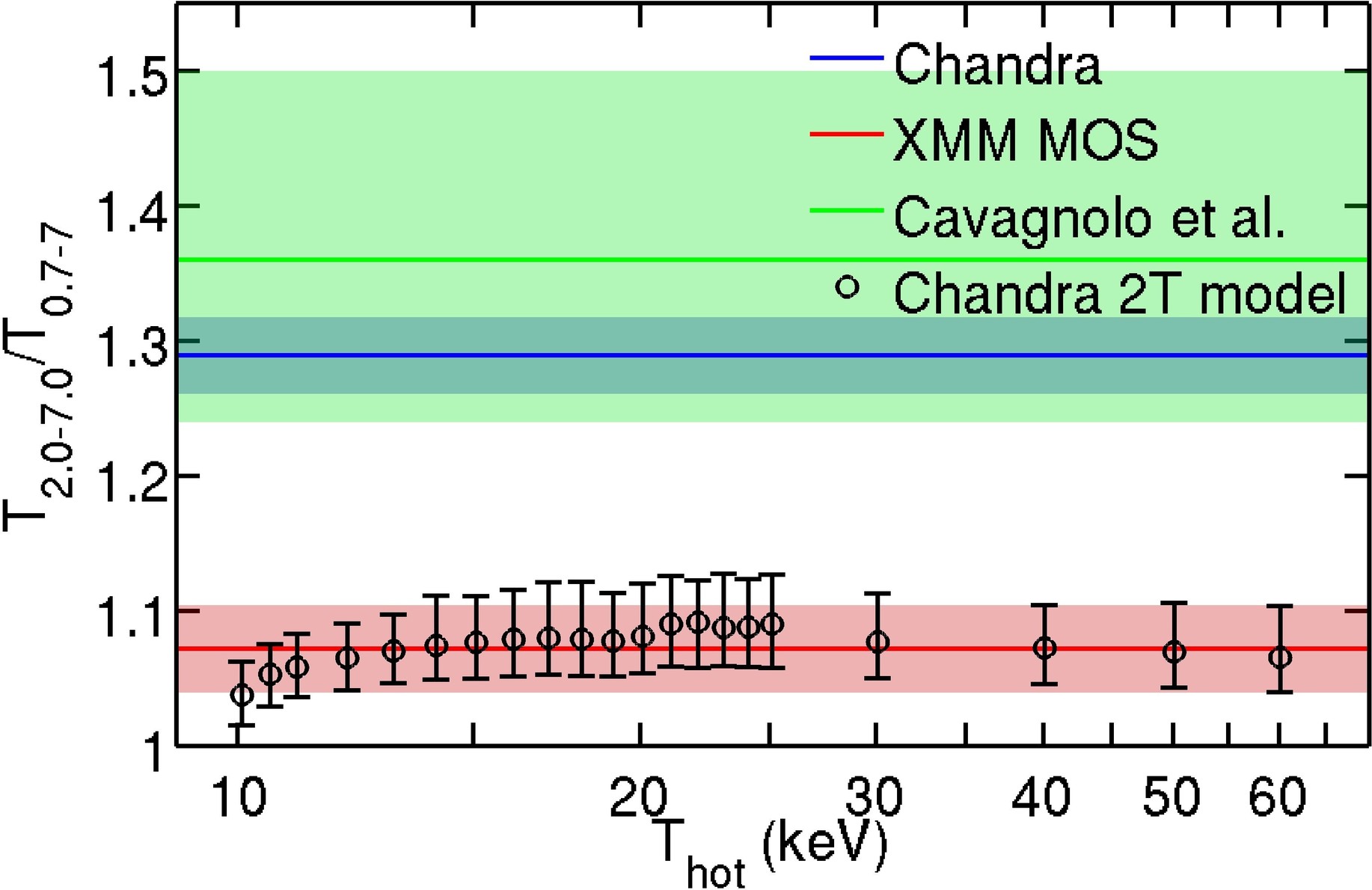}
\caption{The hard-band to broad-band temperature ratio $T_{2.0-7.0}/T_{0.7-7.0}$ of simulated {\it Chandra} 2T spectra (circles) plotted against $\Th$. The shaded regions show the observed temperature ratios from {\it Chandra} and {\it XMM-Newton} MOS data. Also shown is the temperature ratio from 40 ks {\it Chandra} data by \cite{cav08}. \label{plt_Thbr}}
\end{center}
\end{figure}

In addition to multiple-component modeling, measuring the temperature in different band-pass is another way to demonstrate the presence of multiple components. \cite{cav08} reported a very high hard-band to broad-band temperature ratio for A1689, $1.36^{+0.14}_{-0.12}$, from analysis of 40 ks of {\it Chandra} data, suggesting that this could relate to ongoing or recent mergers. Following the convention in \cite{cav08}, we fit the spectrum in the 0.7-7.0 keV (broad) and 2.0/(1+$z$)-7.0 keV (hard) band with a single-temperature model. In contrast to C08, we do not use the $r<R_{2500}$ region with the core excised, but simply take the spectrum from the whole central 3\arcmin~(395 \kpc) region. The hard-band to broad-band temperature ratio from {\it Chandra} data, $1.29\pm0.03$, strongly disagrees with that of {\it XMM} MOS, $1.07\pm0.03$. This result is anticipated since an absorption edge feature found in the {\it Chandra} spectrum (\S \ref{system}) is close to the cut-off of the hard band. After correcting for this absorption, the temperature ratio is in the range of 1 to 1.08 for an absorption depth of $\tau=0.14-0.10$. As a consistency check, we simulated spectra according to the best-fit 2T models (from {\it Chandra} data) from \S \ref{2T} to see whether these models can explain such a high temperature ratio. Results are plotted in Fig.\,\ref{plt_Thbr}. None of the 2T models can reproduce the observed ratio of the uncorrected {\it Chandra} data. Thus we conclude that there is no evidence from this ratio of the presence of multiple components or merging activity. Furthermore, \cite{lec08} do not find any discrepancy between the hard band (2-10 keV) and broad band (0.7-10 keV) temperature profiles, except for $r<0.05~r_{180}$, for a sample of $\sim$ 50 hot, intermediate redshift clusters based on {\it XMM-Newton} observations. The high hard-band to broad-band temperature ratio seen in A1689, as well as in many other clusters observed with {\it Chandra} \citep{cav08}, might be due to the aforementioned calibration uncertainty.

\subsection{Emission line diagnostics}\label{line}
\begin{figure}
\includegraphics[width=8.5cm]{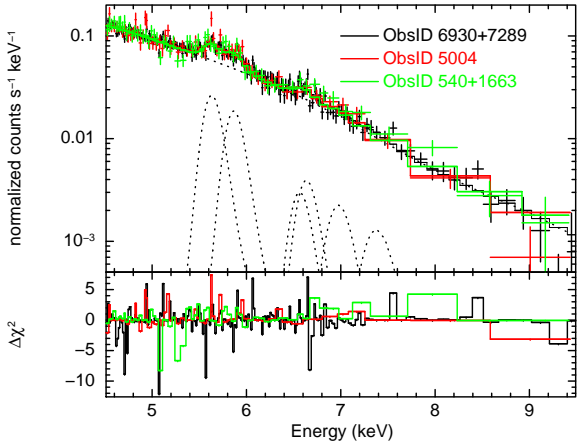}
\caption{The 4.5-9.5 keV {\it Chandra} spectrum of the central 3\arcmin~region. The spectrum is modeled with an absorbed thermal bremsstrahlung plus the seven Gaussian lines listed in Table\,\ref{line}. 
\label{plt_specline}}
\end{figure}

When fitting the whole spectrum, the temperature is mainly determined by the continuum due to the low amount of line emission at the temperature of A1689. In order to extract the emission line information, which can provide an additional temperature diagnostic, we fit the 4.5-9.5 keV spectrum with an absorbed thermal bremsstrahlung model plus Gaussians. There are 42 lines whose emissivity is greater than $10^{-19}$ photons cm$^{-3}$s$^{-1}$ at $kT=10~$keV from ions of Fe \textsc{xxv}, Fe \textsc{xxvi}, Ni \textsc{xxvii}, Ni \textsc{xxviii}, according to {\it Chandra} ATOMDB 1.3.1. Considering the CCD energy resolution, we grouped those lines into seven Gaussians and used the emissivity-weighted centroid and one standard deviation as the line center and width, respectively. The Ni \textsc{xxvii} K$\alpha$ line is $\sim$ 80 eV away from the Fe \textsc{xxv} K$\beta$ line, not separable under CCD resolution unless we have extremely good data quality. Since we obtained an unusually high Ni/Fe ratio of $\sim$ 9 Ni$_{\sun}$/Fe$_{\sun}$ from a VAPEC model fit to the whole spectrum (\S \ref{1T}), it is worth to investigate this in detail. We therefore modeled Ni \textsc{xxvii} K$\alpha$ and Fe \textsc{xxv} K$\beta$ lines individually. Fig.\,\ref{plt_specline} shows the spectrum and the best-fit model. The modeled lines are listed in Table\,\ref{tbl_line}.

\begin{deluxetable}{llll}
\tablewidth{0pt}
\tabletypesize{\footnotesize}
\tablecaption{\label{tbl_line}}
\tablehead{
\colhead{Line} & 
\colhead{Energy} & 
\colhead{Centroid\tna} &  
\colhead{Width\tna} \\
     &  \colhead{(keV)}  &  \colhead{(keV)}   & \colhead{(eV)}}
\startdata
Fe \textsc{xxv} K$\alpha$   & 6.636, 6.668, 6.682, 6.700 & 6.686 & 23  \\
Fe \textsc{xxv} K$\beta$    & 7.881                      & 7.877 & 19  \\
Fe 8.3 keV\tnb              & 8.246, 8.252, 8.293, 8.486 & 8.282 & 68  \\
Fe \textsc{xxvi} K$\alpha$  & 6.952, 6.973               & 6.964 & 14  \\
Fe 8.7 keV\tnc              & 8.698, 8.701, 8.907, 8.909 & 8.764 & 97  \\
Ni \textsc{xxvii} K$\alpha$ & 7.765, 7.805               & 7.793 & 19  \\
Ni \textsc{xxviii} K$\alpha$& 8.074, 8.101               & 8.090 & 16  \\[-7pt]
\enddata
\tablenotetext{a}{Emissivity-weighted center and one standard deviation. The line emissivity is calculated at $T=10$ keV from {\it Chandra} ATOMDB 1.3.1.}
\tablenotetext{b}{including Fe \textsc{xxvi} K$\beta$, \textsc{xxv} K$\gamma$, and \textsc{xxv} K$\delta$.}
\tablenotetext{c}{including Fe \textsc{xxvi} K$\delta$ and \textsc{xxvi} K$\gamma$.}
\end{deluxetable}

\begin{figure*}
\begin{center}
\includegraphics[width=10cm,angle=270]{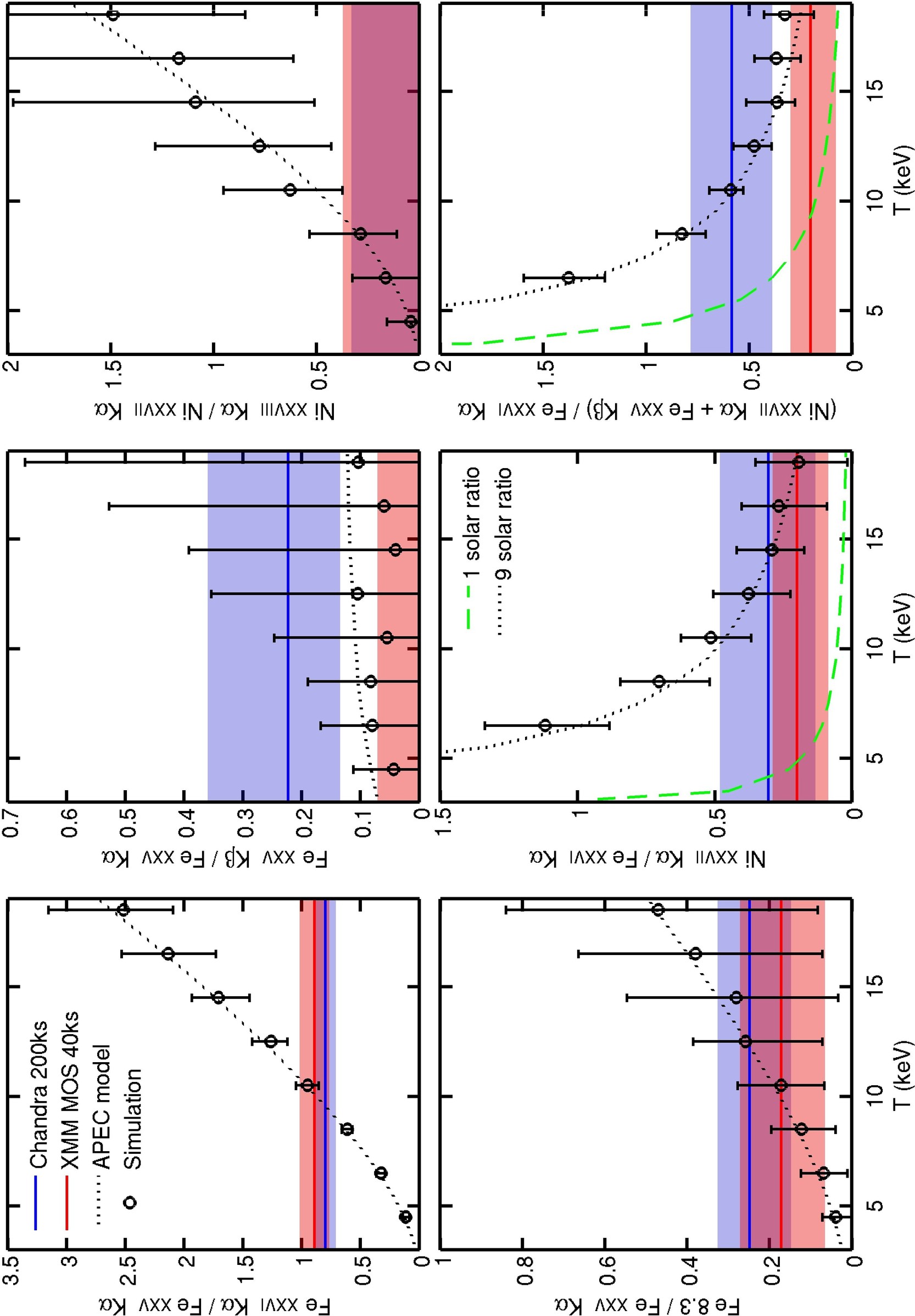}
\caption{The predicted 1T plasma line ratio (dotted line) as a function of temperature, for various lines. The observed ratio and its 1$\sigma$ confidence are shown as a solid line and shaded region. The circles show the fitted results of 100 simulated {\it Chandra} spectra drawn from a VAPEC model with 9 Ni$_{\sun}$/Fe$_{\sun}$. \label{plt_lineratio}}
\end{center}
\end{figure*}

Strictly speaking, using fixed values of line centroids and widths is not correct because those quantities change with temperature. In addition, we approximated the line complex as a Gaussian whose line centroid and width calculated from the model may not be the same after being convolved with the instrument response. To properly compare our fit results with the theory, we simulated spectra and fit them the same way we fit the real data. Fig.\,\ref{plt_lineratio} shows the observed line ratios and results from simulated VAPEC spectra with 9 Ni$_{\sun}$/Fe$_{\sun}$. 100 spectra were produced at each temperature and the flux was kept at the same level as that of the data. From the good match of fitted results from simulations to the direct model prediction, we confirmed that the fitting is accurate enough to measure the line flux, though only Fe \textsc{xxv} K$\alpha$ and Fe \textsc{xxvi} K$\alpha$ lines are precise enough for temperature determination. Table\,\ref{tbl_lineT} shows the temperature and abundances, inferred from a single-temperature APEC model. The iron line temperature is in very good agreement with the continuum temperature for both {\it Chandra} and {\it XMM-Newton} data. All the {\it Chandra} and {\it XMM-Newton} observed line fluxes, except Fe \textsc{xxv} K$\beta$, are consistent with each other (after an overall 9\% adjustment to the flux). Using Fe \textsc{xxv}+\textsc{xxvi} K$\alpha$ and Ni \textsc{xxvii} K$\alpha$ line flux, we obtain accordant Fe and Ni abundances from both instruments. The larger Fe and Ni abundances found in \S \ref{1T} for {\it Chandra} data are likely due to the higher temperature determined by the broad-band spectrum and the much stronger Fe \textsc{xxv} K$\beta$ line.

\begin{deluxetable}{llll}
\tablewidth{0pt}
\tabletypesize{\footnotesize}
\tablecolumns{4}
\tablecaption{Summary of line analysis \label{tbl_lineT}}
\tablehead{         &            &  \colhead{{\it Chandra}}   &  \colhead{{\it XMM} MOS} }
\startdata
\sidehead{Continuum}
$T$                     &(keV)                & $10.3^{+ 2.2}_{-0.8}$& $ 9.7^{+ 0.8}_{-1.1}$  \\
\tableline
\sidehead{Emission lines}
$T$\tna    &(keV)                & $9.6^{+0.5}_{-0.5}$    & $10.1^{+0.7}_{-0.7}$  \\
Ni/Fe\tnb\tndg &(Ni$_\Sun$/Fe$_\Sun$)& $8.4^{+3.7}_{-3.6}$    & $1.4^{+1.8}_{-1.4}$   \\
Ni/Fe\tnc\tndg &(Ni$_\Sun$/Fe$_\Sun$)& $5.5^{+3.2}_{-3.1}$    & $3.7^{+1.6}_{-2.1}$   \\
\tableline
Fe\tnd\tndg    &(Z$_\Sun$)           & $0.31\pm0.02$          & $0.32\pm0.03$         \\
Ni\tne\tndg    &(Z$_\Sun$)           &$1.23^{+0.50}_{-0.91}$  & $1.08^{+0.52}_{-0.65}$\\[-7pt]
\enddata
\tablenotetext{a}{from Fe \textsc{xxvi} K$\alpha$/Fe \textsc{xxv} K$\alpha$.}
\tablenotetext{b}{from (Ni \textsc{xxvii} K$\alpha$+Fe \textsc{xxv} K$\beta$)/Fe \textsc{xxvi} K$\alpha$.}
\tablenotetext{c}{from Ni \textsc{xxvii} K$\alpha$/Fe \textsc{xxvi} K$\alpha$.}
\tablenotetext{d}{from (Fe \textsc{xxvi} K$\alpha$+\textsc{xxv} K$\alpha$)/continuum.}
\tablenotetext{e}{from Ni \textsc{xxvii} K$\alpha$/continuum.}
\tablenotetext{$\dagger$}{assuming $T=10$ keV.}
\end{deluxetable}

\begin{figure*}
\begin{center}
\includegraphics[width=5cm,angle=270]{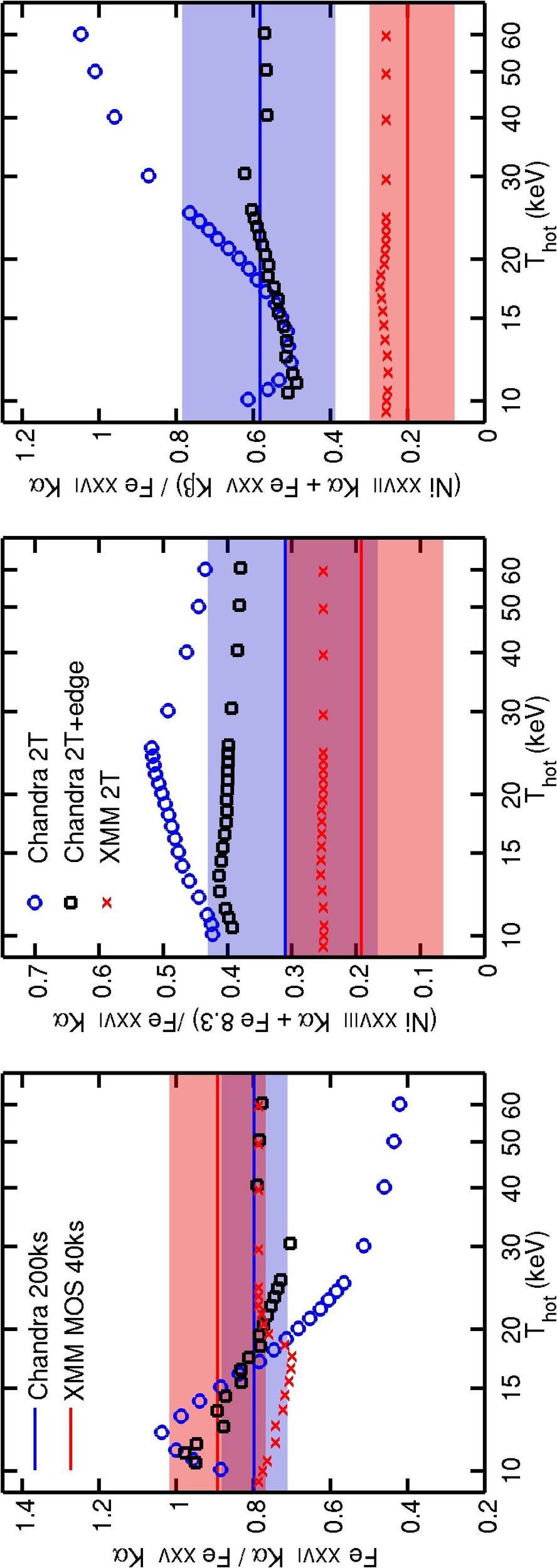}
\caption{The predicted line ratio from the best-fit 2T (VAPEC) models (\S \ref{2T}) as a function of the temperature of the hotter phase $\Th$. The solid line and shaded region shows the observed ratio and its 1$\sigma$ error. The x-axis is in log scale.\label{plt_lineratio2}}
\end{center}
\end{figure*}

As discussed previously, the 2T analysis of {\it Chandra} data suggested that another spectral component is needed if no absorption edge modeling is applied. Fig.\,\ref{plt_lineratio2} shows the line ratios predicted by the best-fit models from \S \ref{2T} over a wide range in temperature for the hot phase. Since the spectral energy range used in this fitting is far enough from the Si edge, it is not necessary to modify the spectral model even if the Si edge indeed needs to be corrected. The broad-band {\it Chandra} spectrum is not sensitive to the hot phase temperature of the 2T model once it exceeds 15 keV (Fig.\,\ref{plt_2t_chi}). With the good constraint from the Fe \textsc{xxvi} K$\alpha$/Fe \textsc{xxv} K$\alpha$ line ratio, models with $\Th>20$ keV, which are composed of great amounts of cooler gas, are rejected.  Meanwhile, the ratio of higher energy states (Ni \textsc{xxviii} K$\alpha$, Fe \textsc{xxvi} K$\beta$, Fe \textsc{xxv} K$\gamma$, K$\delta$) to the well-measured Fe \textsc{xxvi} K$\alpha$ line suggests that models with lower $\Th$ are preferable.

As for the 2T models based on {\it Chandra} with an absorption edge model and {\it XMM-Newton} broad-band spectra, predicted line ratios all agree with the observed value. In fact, models with $\Th>20$ keV from {\it XMM-Newton} data are essentially a one temperature model, since the normalization of the hot component in these models is zero. Adding the fact that an additional temperature component does not significantly improve the $\chi^2$ of the fit for those spectra and the remarkably good agreement on the temperature measured by the continuum and the iron lines from both {\it Chandra} and {\it XMM-Newton}, we conclude that the simple 1T model is adequate to describe the X-ray emission from the central 3\arcmin~ region of A1689.

\section{DEPROJECTION ANALYSIS}\label{sec_proj}
Assuming that the hotter phase gas has the 3D temperature profile of \citetalias{lem07}, the radial distribution of the cooler gas can be derived. We extracted spectra from concentric annuli up to 8.8\arcmin~(1.2 \Mpc). The emission from each shell in three-dimensional space was modeled with an absorbed two-temperature APEC model with $\Th$ fixed at the value of \citetalias{lem07} and then projected by the PROJCT model in XSPEC. Because of the complexity of this model, we used coarser annular bins than those used in \citetalias{lem07}. Data of \citetalias{lem07} were binned using the weighting scheme of \cite{maz04} to produce a spectroscopic-like temperature. $\Tc$, abundance, and the normalization of both components were free to vary. The outermost two annuli were background dominated, so spectra were binned to have at least 15 net counts per bin at $r=$ 4.8\arcmin-6.5\arcmin~(625-852 \kpc) and 2 net counts at $r=$ 6.5\arcmin-8.8\arcmin~(852-1161 \kpc) (see \S \ref{fitting}). \citetalias{lem07} predicted the gas temperature only up to 721 \kpc, and that temperature was slightly below the observed one. Therefore, we allowed $\Th$ to change in the last two bins. The cold component was removed and the abundance was fixed at 0.2 solar in these regions in order to constrain the rest of the parameters better.

\begin{figure*}
\begin{center}
\includegraphics[width=8.2cm]{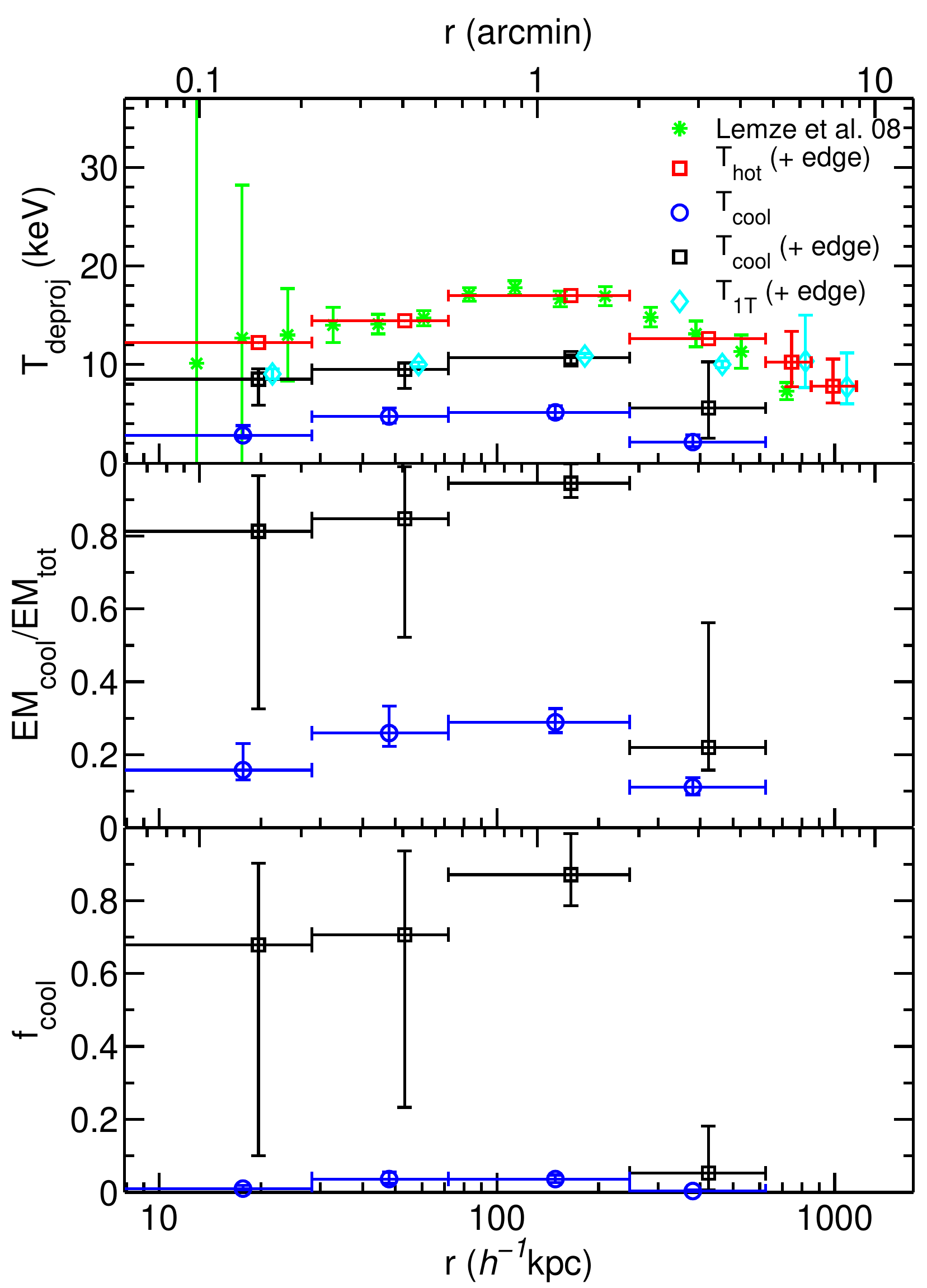}\includegraphics[width=8.5cm]{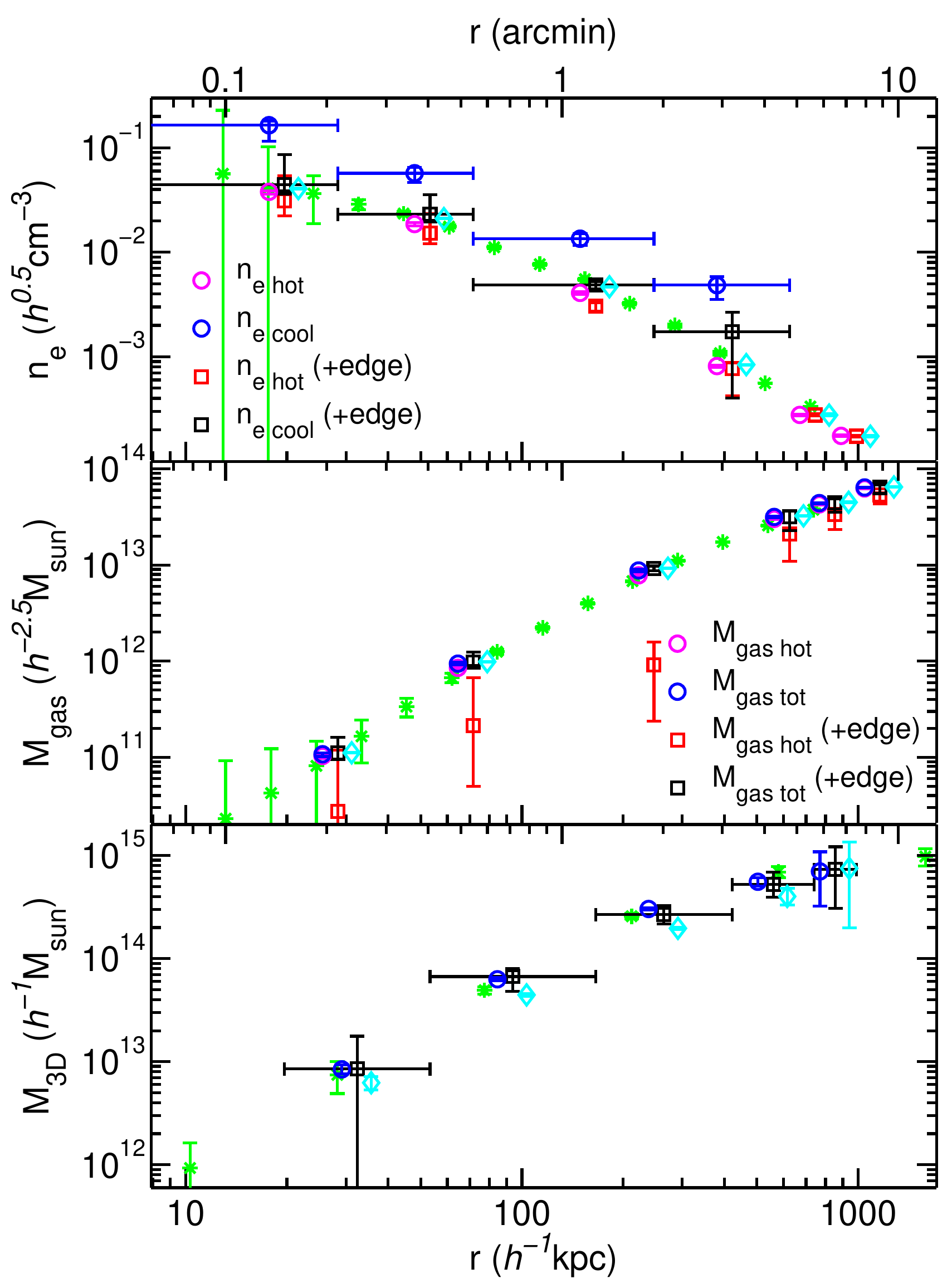}
\caption{Temperature, emission measure ratio of the cool component $EM_{\text{cool}}/EM_{\text{tot}}$, volume filling fraction of the cool component $f_{\text{cool}}$, gas number density $n_e$, cumulative gas mass $M_{\text{gas}}$, and cumulative total mass $M_{3D}$ profiles from the 2T deprojection analysis with an absorption edge correction (squares) and without the correction (circles). Also shown is the 1T analysis (diamonds) and results from \cite{lem07} (asterisks). $\Th$ of the first 4 annuli was fixed at the value derived from lensing and X-ray brightness data \citep{lem07}, which were grouped into fewer bins. The cool component of the last 2 bins was frozen at zero. The 2T assumption is held within 625 \kpc. X-data points of the 1T and 2T models have been shifted by +10\% and -10\% for clarity, and their error bars are also omitted. \label{plt_proj}}
\end{center}
\end{figure*}

Assuming two phases in pressure equilibrium, the volume filling fraction of the $i$th component can be obtained from
\begin{equation}
f_i=\frac{\mbox{Norm}_iT_i^2}{\sum_{j}\mbox{Norm}_jT_j^2}
\end{equation}
\citep[e.g.,][]{san02}. Once $f_i$ is determined, the gas density $\rho_{gi}=\mu_e m_p n_{ei}$ can be derived from 
\begin{equation}
\mbox{Norm}_i = \frac{10^{-14}}{4\pi((1+z)D_A)^2}\int n_{ei}n_{Hi}f_idV,
\end{equation}
where $n_H$/$n_e$ and $\mu_e$ are calculated from a fully ionized plasma with the measured abundance \citep[He abundance is primordial, and others are from][]{and89}. For $Z=0.3$ \Zsun, $n_H$/$n_e=0.852$ and $\mu_e=1.146$. Fig.\,\ref{plt_proj} shows the results of this deprojected 2T analysis. The 1T modeling, in which emission from each shell has only one component, and the results from \citetalias{lem07}, are also shown. If the cluster has a temperature profile of \citetalias{lem07}, then 70-90\% of the space within 250 \kpc~is occupied by the "cool" component with a temperature of $\sim$ 10 keV, based on {\it Chandra} absorption edge corrected data, and this gas constitutes 90\% of the total gas mass. 

\cite{kaw07} show that local density and temperature inhomogeneities do not correlate with each other in simulated clusters, which undermines the assumption of two phases in thermal pressure equilibrium. However, other cosmological simulations find that gas motions contribute about 5-20\% of the total pressure support \citep[e.g.,][]{fal05,ras06,lau09}. If the pressure balance is off by 20\%, it will not significantly change the gas mass fraction ($\lesssim4$\%) or the volume filling fraction ($\lesssim8$\%).

\section{MASS PROFILE}\label{sec_mass}
Given the 3D gas density and temperature profiles, the total cluster mass within a radius $r$ can be estimated from the hydrostatic equilibrium equation \citep[e.g.,][]{sar88}, 
\begin{equation}
M(r)=- \frac{kT(r)\,r}{G\mu\,m_{p}} \left( \frac{d\ln \rho_{g}(r)}{d\ln r}+ \frac{d\ln T(r)}{d\ln r} \right), 
\label{eq_mass}
\end{equation}
For $Z=0.3$ \Zsun, $\mu=0.596$.

If the gas has two temperatures with two phases in pressure equilibrium, the total mass still can be derived from Eq.~\ref{eq_mass} with $\rho_g$, $T$ replaced by $\rhogh$, $\Th$, respectively.

\subsection{Nonparametric method}\label{nonparfit}
To evaluate the derivatives in Eq.~\ref{eq_mass}, we took the differences of deprojected temperature and the gas density in log space. The radius of each annulus was assigned at $\bar{r}$ such that 
\begin{equation}
F_{3D}(\bar{r})\,\frac{4\pi}{3}\left( r_{out}^3-r_{in}^3\right)=\int^{r_{out}}_{r_{in}} F_{3D}(r)\,4\pi r^2dr,
\end{equation}
where $F_{3D}$ is the deprojected flux density from a finely binned surface brightness profile, and $r_{in}$ ($r_{out}$) is the inner (outer) radius of the annulus. The radius $r$ outside of the brackets of Eq.~\ref{eq_mass} is taken at the geometric mean (i.e. the arithmetic mean in log scale) of the radii of two adjacent rings, $r=\sqrt{{\bar{r}_i\,\bar{r}_{i+1}}}$ , and the temperature is linearly interpolated at this radius. Because errors from e.g. $T$ and $dT/dr$ are not independent, standard error propagation is not easily applied. Uncertainties are estimated from the distribution of 1000 Monte-Carlo simulations of $T$ and $\rho_g$ profiles. Fig.\,\ref{plt_proj} shows the total mass profile from both 1T and 2T models and the results are listed in Table\,\ref{tbl_mass}. Two-temperature modeling, based on the $T_{hot}$ of L08, increases the total mass by 30-50\% for all radii within 625 \kpc. Beyond that radius, the 2T assumption is not held because of the lack of constraint on $\Th$.

\begin{deluxetable}{cccc}
\tablewidth{0pt}
\tabletypesize{\footnotesize}
\tablecaption{Total mass profile\label{tbl_mass}}
\tablehead{
\colhead{$r$} & 
\colhead{$\MTe$} & 
\colhead{$\MTTe$} &  
\colhead{$\MTT$}\\
\colhead{(\kpc)} & 
\colhead{($10^{14}$\hMsun)} & 
\colhead{($10^{14}$\hMsun)} &  
\colhead{($10^{14}$\hMsun)}}
\startdata
 $  32^{+  21}_{ -13}$ & $0.06^{+0.01}_{-0.01}$ & $0.09^{+0.09}_{-0.08}$ & $0.08^{+0.01}_{-0.01}$  \\
 $  94^{+  72}_{ -41}$ & $0.44^{+0.01}_{-0.01}$ & $0.67^{+0.13}_{-0.19}$ & $0.63^{+0.02}_{-0.02}$  \\
 $ 264^{+ 158}_{ -99}$ & $1.96^{+0.04}_{-0.05}$ & $2.67^{+0.60}_{-0.50}$ & $3.03^{+0.06}_{-0.07}$  \\
 $ 559^{+ 181}_{-137}$ & $4.01^{+0.80}_{-0.69}$ & $5.23^{+1.67}_{-1.31}$ & $5.57^{+0.53}_{-0.50}$  \\
 $ 855^{+ 133}_{-115}$ & $7.50^{+6.04}_{-5.51}$ & $7.33^{+4.84}_{-4.27}$ & $7.00^{+3.88}_{-3.76}$  \\[-7pt]
\enddata
\tablecomments{2T assumption is only held within 625 \kpc. The upper and lower limits of $r$ indicate the radii $\bar{r}$ of two contiguous rings used to calculate the mass. See text for definitions of $r$ and $\bar{r}$.}
\end{deluxetable}

Although the inclusion of an absorption edge in the spectral model greatly changes the derived composition of the multi-phase plasma, it does not affect the mass measurement much. This is because we use a fixed $\Th$ profile. Once the temperature is determined, the total mass only depends on the logarithmic scale of the gas density, which produces $\sim$ 13\% difference at most. 

\subsection{1T parametric method}\label{parfit}
If the temperature does not vary dramatically on small scales, we can obtain a mass profile with higher spatial resolution since the gas density can be measured in detail from the X-ray surface brightness with the assumption of a certain geometry of the cluster. To achieve this, modeling of the temperature and the gas density is necessary. Following the procedure of \cite{vik06}, we project the 3D temperature and the gas density models along the line of sight and fit with the observed projected temperature and the surface brightness profiles. A weighting method by \cite{maz04}, \cite{vik06spec} is used to predict a single-temperature fit to the projected multi-temperature emission from 3D space. This method has been shown \citep{nag07} to accurately reproduce density and temperature profiles of simulated clusters. 

\begin{figure}
\begin{center}
\includegraphics[width=8.5cm]{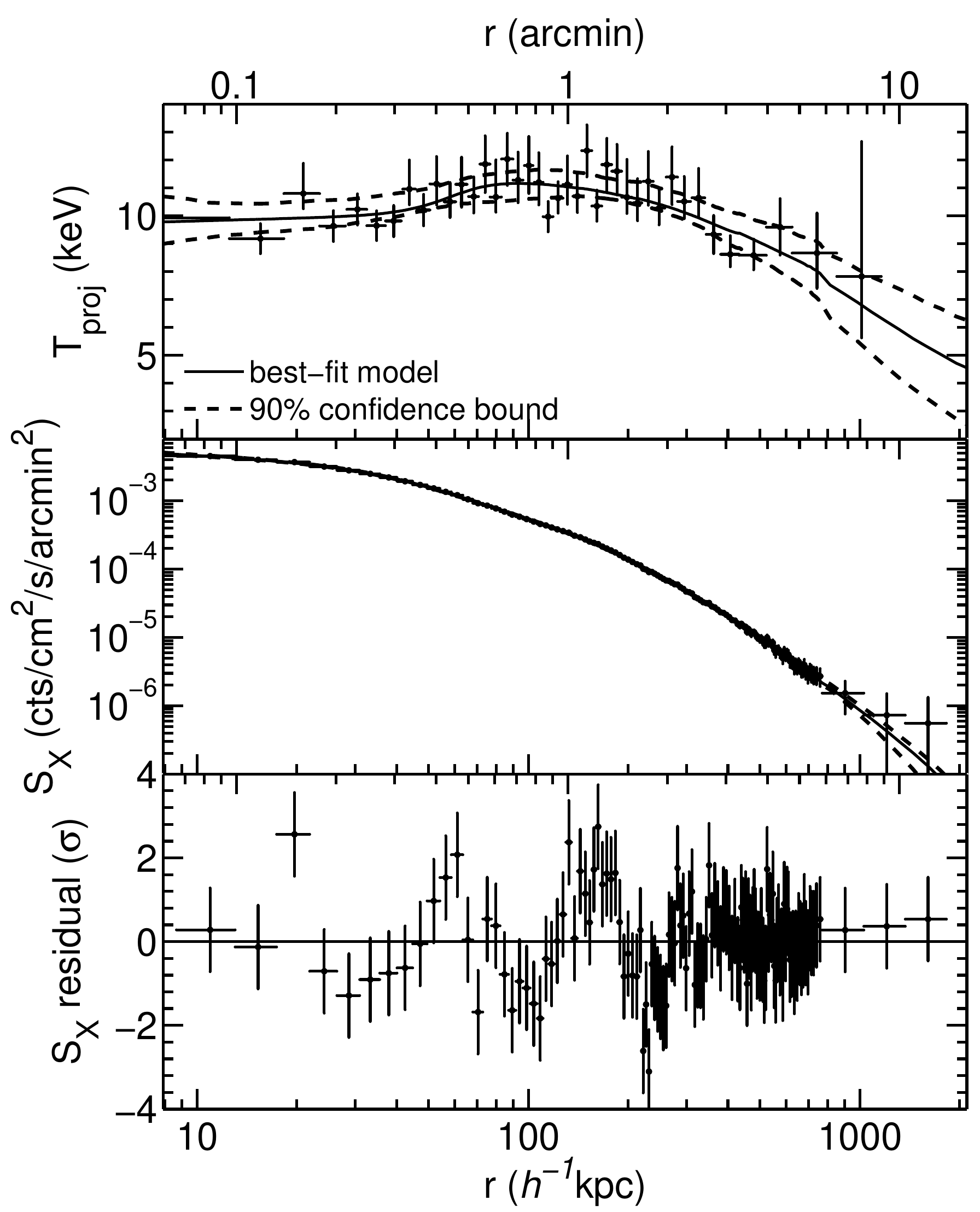}
\caption{Projected temperature and surface brightness profiles with the best-fit model (solid lines) and its 90\% confidence bounds (dashed lines). Bottom panel: residual between the surface brightness and the model. This fit gives a $\chi^2/dof$ of 154.3/155.
\label{plt_ts}}
\end{center}
\end{figure}

\begin{figure}
\begin{center}
\includegraphics[width=8.5cm]{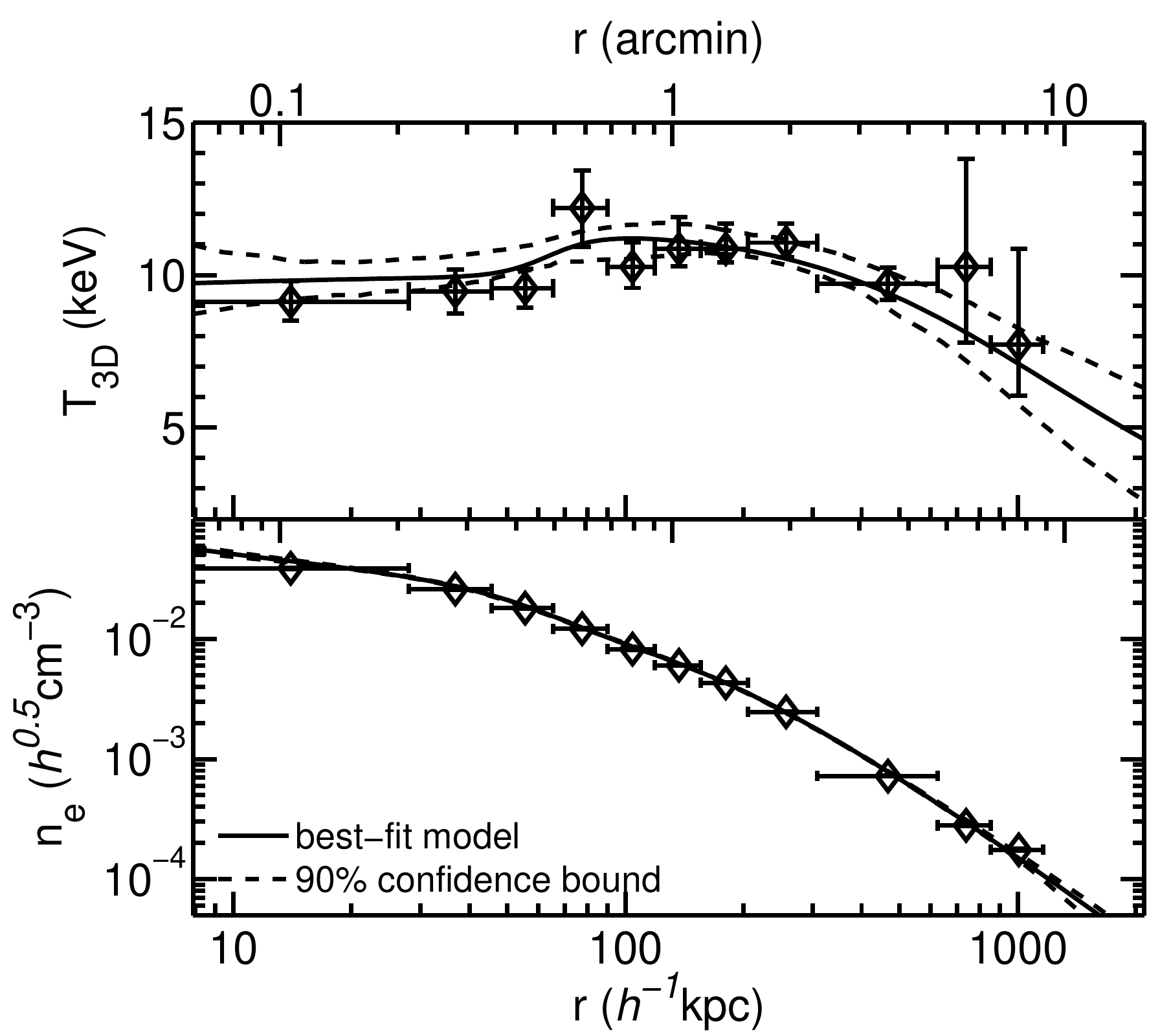}
\caption{Best-fit $T_{\text{3D}}$ and $n_e$ models (solid lines) and 90\% confidence bounds (dashed lines). Also shown are unparameterized results (diamonds) from \S \ref{nonparfit}.
\label{plt_tn3d}}
\end{center}
\end{figure}

 The gas density model is given by 
\begin{equation}\label{eq_density}
  \begin{split}
    n_p\,n_e = n_0^2\;\frac{(r/r_c)^{-\alpha}}{(1+r^2/r_c^2)^{3\beta-\alpha/2}}
                \;&
                \frac{1}{(1+r^\gamma/r_s{}^\gamma)^{\varepsilon/\gamma}}\\
                + \frac{n_{02}^2}{(1+r^2/r_{c2}^2)^{3\beta_2}},  
  \end{split}
\end{equation}
which originates from a $\beta$ model \citep{cav78} modified by a power-law cusp and a steepening at large radii \citep{vik99}. The second term describes a possible component in the center, especially for clusters with small core radius. The temperature model is given by
\begin{equation}\label{eq_temp}
   T_{\text{3D}}(r) =  T_0\frac{(r/r_{\text{cool}})^{a_{\text{cool}}}+T_{\text{min}}/T_0}{1+(r/r_{\text{cool}})^{a_{\text{cool}}}} \frac{(r/r_t)^{-a}}{(1+(r/r_t)^b)^{c/b}}, \\
\end{equation}
which is a broken power law with central cooling \citep{all01}. Best-fit parameters for the gas density and temperature profiles are listed in Tables \ref{tbl_density} and \ref{tbl_temp}, respectively. Errors are estimated from the distribution of the fitted parameters of 1000 simulated projected temperature and surface brightness profiles generated according to the observed data and their measurement uncertainties. Since parameters are highly degenerate, some of the best-fit values are not covered by the upper or lower limits with the quoted confidence level (upper/lower bounds are for one parameter). The observed temperature and surface brightness profiles, the best-fit model, and the surface brightness residual are shown in Fig.\,\ref{plt_ts}. The model describes the data very well ($\chi^2/dof$=154.3/155). The best-fit $T_{\text{3D}}$ and $n_e$ models are shown in Fig.\,\ref{plt_tn3d}. Also plotted are the profiles from the spectral deprojection fitting (\S \ref{nonparfit}).  Compared to this nonparametric result, modeling $T_{\text{3D}}$ and $n_e$ can avoid flucutations from the direct spectral deprojection, which is a common problem as the deprojection tends to amplify the noise in the data \citep[see Appendix in][]{san07}.

Although the second break of the first term in Eq.~\ref{eq_density} was designed to describe the steepening at $r_s > 0.3\,r_{200}$ \citep{vik99,vik06,neu05}, we found that if the initial guess for $r_s$ is not big enough, $r_s$ tends to converge to a relatively small value, $\approx 200$ \kpc, compared to the typical value of 400-3000 \kpc~for nearby relaxed clusters \citep{vik06}.  It is possible to use the first core radius $r_c$ or the core radius of the second component $r_{c2}$ to account for the sharpening at 200 \kpc. This consequently yields a more reasonable $r_s$ at $\approx 1$ \Mpc. Both cases, small (Model 1) and large (Model 2) $r_s$, give acceptable fits with $\chi^2/dof$ of 153.4/155 and 154.3/155, respectively. However, large $r_s$ is harder to constrain. This makes the mass estimate more uncertain at large radii than the small $r_s$ case. 

Comparing the surface brightness profile of the northeastern (NE) part to the southwestern (SW), \cite{rie08} found that the NE part is 5-15\% brighter outside 350 \kpc~and 25\% under-luminous at 70 \kpc~than the SW. To see if this asymmetry can affect the mass estimate, we fit a symmetric model to the image and iteratively removed any part of the cluster that deviates significantly from the azimuthal mean, mainly the northern clump at 460 \kpc, the southern less luminous region at 330 \kpc, and possibly some point sources not completely removed beforehand. We did not exclude these regions from our temperature measurement since they were unlikely to bias the average temperature much for such a hot cluster, as shown in Fig.\,\ref{plt_2t_chi} that at least 10-20\% of the total emission measure from another spectral component was needed in order the change the spectroscopic temperature by 1 keV. Best-fit gas density and temperature for these models are listed in Tables \ref{tbl_density} and \ref{tbl_temp}, labelled with Model 3 (small $r_s$) and 4 (large $r_s$). 

\begin{deluxetable*}{lcccccccccc}
\addtolength{\tabcolsep}{-4pt}
\tabletypesize{\scriptsize}
\tablecaption{Best-fit parameters for the gas density (Eq.\ref{eq_density})\label{tbl_density}}
\tablehead{
\colhead{} &
\colhead{$n_0$} &
\colhead{$r_c$} &
\colhead{$r_s$} &
\colhead{$\alpha$} &
\colhead{$\beta$} &
\colhead{$\varepsilon$} &
\colhead{$n_{02}$} &
\colhead{$r_{c2}$} &
\colhead{$\beta_2$} &
\colhead{$\gamma$} \\
\colhead{} &
\colhead{$10^{-2}h^{\frac{1}{2}}$cm$^{-3}$} &
\colhead{$10^2$\kpc} &
\colhead{$10^2$\kpc} &
\colhead{} &
\colhead{} &
\colhead{} &
\colhead{$10^{-1}h^{\frac{1}{2}}$cm$^{-3}$} &
\colhead{$10^2$\kpc} &
\colhead{} &
\colhead{}
 }
\startdata
(1) & $ 3.68^{+ 0.01}_{-0.78}$ &$ 0.25^{+ 0.10}_{-0.00}$ &$ 1.99^{+ 0.19}_{-0.10}$ &$ 0.37^{+ 0.32}_{-0.13}$ &$ 0.36^{+ 0.04}_{-0.03}$ &$ 1.89^{+ 0.24}_{-0.28}$ &$ 0.20^{+ 0.03}_{-0.03}$ &$ 2.04^{+ 0.26}_{-0.49}$ &$ 7.10^{+ 2.54}_{-2.63}$ &$ 4.64^{+ 1.23}_{-0.99}$ \\
(2) & $ 0.78^{+ 0.14}_{-0.07}$ &$ 1.65^{+ 0.11}_{-0.21}$ &$11.5^{+17.4}_{+0.1}$ &$ 1.14^{+ 0.08}_{-0.16}$ &$ 0.73^{+ 0.02}_{-0.04}$ &$ 0.50^{+ 3.66}_{-0.48}$ &$ 0.27^{+ 0.03}_{-0.01}$ &$ 1.72^{+ 0.16}_{-0.72}$ &$ 6.10^{+ 0.85}_{-3.65}$ &$ 2.56^{+ 7.06}_{-0.28}$ \\
(3) & $ 2.46^{+ 1.08}_{-0.22}$ &$ 0.46^{+ 0.03}_{-0.15}$ &$ 2.70^{+ 0.19}_{-0.15}$ &$ 0.73^{+ 0.09}_{-0.39}$ &$ 0.44^{+ 0.02}_{-0.05}$ &$ 1.91^{+ 0.36}_{-0.09}$ &$ 0.19^{+ 0.02}_{-0.04}$ &$ 2.33^{+ 0.14}_{-0.89}$ &$10.0^{+ 0.0}_{-6.1}$\tna &$ 7.52^{+ 0.46}_{-2.56}$ \\
(4) & $ 0.64^{+ 0.33}_{-0.19}$ &$ 1.98^{+ 0.80}_{-0.60}$ &$10.2^{+27.9}_{-6.72}$ &$ 1.20^{+ 0.11}_{-0.56}$ &$ 0.77^{+ 0.19}_{-0.12}$ &$ 1.45^{+ 2.91}_{-1.33}$ &$ 0.27^{+ 0.09}_{-0.02}$ &$ 1.77^{-0.19}_{-1.22}$ &$ 6.01^{-0.69}_{-5.08}$ &$ 2.08^{+ 8.52}_{-0.71}$ \\[-7pt]
\enddata
\tablecomments{(1) small $r_s$, (2) large $r_s$, (3) small $r_s$ with northern clumps removed, (4) large $r_s$ with northern clumps removed. Errors are 95\% CL for one parameter from 1000 Monte-Carlo simulations. Since parameters are highly degenerate, some of the best-fit values are not covered by the upper and lower limits at this confidence level.}
\tablenotetext{a}{parameters hit the hard limit.}
\end{deluxetable*}

\begin{deluxetable*}{lcccccccccc}
\tablewidth{0pt}
\tabletypesize{\scriptsize}
\tablecaption{Best-fit parameters for the temperature (Eq.\ref{eq_temp})\label{tbl_temp}}
\tablehead{
\colhead{} &
\colhead{$T_0$} &
\colhead{$T_{\text{min}}/T_0$} &
\colhead{$r_{\text{cool}}$} &
\colhead{$r_t$} &
\colhead{$a$} &
\colhead{$b$} &
\colhead{$c$} &
\colhead{$d$} \\
\colhead{} &
\colhead{keV} &
\colhead{} &
\colhead{$10^2$\kpc} &
\colhead{$10^2$\kpc} &
\colhead{} &
\colhead{} &
\colhead{} &
\colhead{}
 }
\startdata
(1) & $12.7^{+ 6.9}_{-2.8}$ &$ 0.73^{+ 0.25}_{-0.34}$ &$ 0.67^{+ 1.66}_{-0.29}$ &$12.5^{+20.8}_{-6.8}$ &$ 0.02^{+ 0.11}_{-0.14}$ &$ 0.86^{+ 0.64}_{-0.36}$ &$ 0.81^{+ 1.82}_{-0.40}$ &$ 2.96^{+ 2.14}_{-2.45}$ \\
(2) & $12.1^{+ 7.2}_{-3.4}$ &$ 0.87^{+ 0.11}_{-0.41}$ &$ 0.64^{+ 1.75}_{-0.15}$ &$ 8.36^{+21.7}_{-0.93}$ &$-0.02^{+ 0.16}_{-0.11}$ &$ 1.37^{+ 0.65}_{-0.80}$ &$ 0.89^{+ 2.11}_{-0.43}$ &$ 6.78^{-0.83}_{-6.30}$ \\
(3) & $14.4^{+ 5.8}_{-5.3}$ & $ 0.40^{+ 0.60}_{-0.13}$ & $ 0.77^{+ 1.42}_{-0.28}$ & $28.6^{+ 0.5}_{-25.0}$ &$ 0.11^{+ 0.10}_{-0.24}$ & $ 0.40^{+ 1.11}_{-0.03}$ & $ 0.59^{+ 1.56}_{-0.19}$ & $ 1.44^{+ 3.86}_{-1.17}$ \\
(4) & $11.8^{+ 8.4}_{-2.9}$ &$ 0.86^{+ 0.14}_{-0.50}$ &$ 0.74^{+ 1.84}_{-0.29}$ &$ 9.08^{+23.1}_{-2.42}$ &$-0.01^{+ 0.17}_{-0.12}$ &$ 1.68^{+ 0.81}_{-1.15}$ &$ 1.32^{+ 1.64}_{-0.92}$ &$ 7.07^{-0.61}_{-6.57}$ \\[-7pt]
\enddata
\tablecomments{(1) small $r_s$, (2) large $r_s$, (3) small $r_s$ with northern clumps removed, (4) large $r_s$ with northern clumps removed. Errors are 95\% CL for one parameter from 1000 Monte-Carlo simulations.}
\end{deluxetable*}

The total mass profiles from these analytic gas density and temperature models are given in Table\,\ref{tbl_parmass}. We list the total mass at the radii where masses from the non-parametric method are evaluated (Table\,\ref{tbl_mass}). The last entry of Table\,\ref{tbl_parmass} shows the total mass at the boundary of the ACIS-I chips, 12\arcmin~(1.6 \Mpc~$\approx r_{200}$), where $S_X$ is detected at $\lesssim 1\sigma$. Removing asymmetric parts from the image or restricting $r_s$ to be greater than 350 \kpc~increases the total mass estimate with $\lesssim$ 10\%. Nonetheless, these differences are not significant. We combine samples of the best-fit parameters of substructures removed cases (Model 3 and 4) as our best result.

\begin{deluxetable}{ccccc}
\tablewidth{0pt}
\tabletypesize{\footnotesize}
\tablecaption{Parametric total mass profile\label{tbl_parmass}}
\tablehead{
\colhead{$r$ (\kpc)} & \multicolumn{4}{c}{$M(r)$ ($10^{14}$\hMsun)} \\
\colhead{} & 
\colhead{(1)} & 
\colhead{(2)} &  
\colhead{(3)} &
\colhead{(4)}
}
\startdata
   32 & $ 0.07^{+ 0.01}_{-0.01}$  & $ 0.07^{+ 0.01}_{-0.01}$ & $ 0.07^{+ 0.01}_{-0.01}$ &  $ 0.07^{+ 0.01}_{-0.01}$ \\
   94 & $ 0.46^{+ 0.01}_{-0.02}$  & $ 0.46^{+ 0.01}_{-0.03}$ & $ 0.47^{+ 0.02}_{-0.02}$ &  $ 0.44^{+ 0.03}_{-0.02}$ \\
  264 & $ 2.01^{+ 0.07}_{-0.03}$  & $ 1.92^{+ 0.09}_{-0.02}$ & $ 1.92^{+ 0.09}_{-0.03}$ &  $ 1.94^{+ 0.08}_{-0.08}$ \\
  559 & $ 4.15^{+ 0.32}_{-0.02}$  & $ 4.43^{+ 0.18}_{-0.11}$ & $ 4.64^{+ 0.27}_{-0.03}$ &  $ 4.86^{+ 0.15}_{-0.51}$ \\
  855 & $ 5.73^{+ 0.46}_{-0.05}$  & $ 6.37^{+ 0.35}_{-0.36}$ & $ 6.41^{+ 0.40}_{-0.23}$ &  $ 7.03^{+ 0.71}_{-1.00}$ \\
 1579 & $ 8.60^{+ 0.59}_{-1.02}$  & $ 9.52^{+ 2.22}_{-0.97}$ & $ 9.85^{+ 0.56}_{-1.52}$ &  $ 9.55^{+ 3.31}_{-1.43}$ \\[-7pt]
\enddata
\tablecomments{(1) small $r_s$, (2) large $r_s$, (3) small $r_s$ with northern clumps removed, (4) large $r_s$ with northern clumps removed. Errors are 68\% CL from 1000 Monte-Carlo simulations.}
\end{deluxetable}

\begin{figure}
\begin{center}
\includegraphics[width=8.5cm]{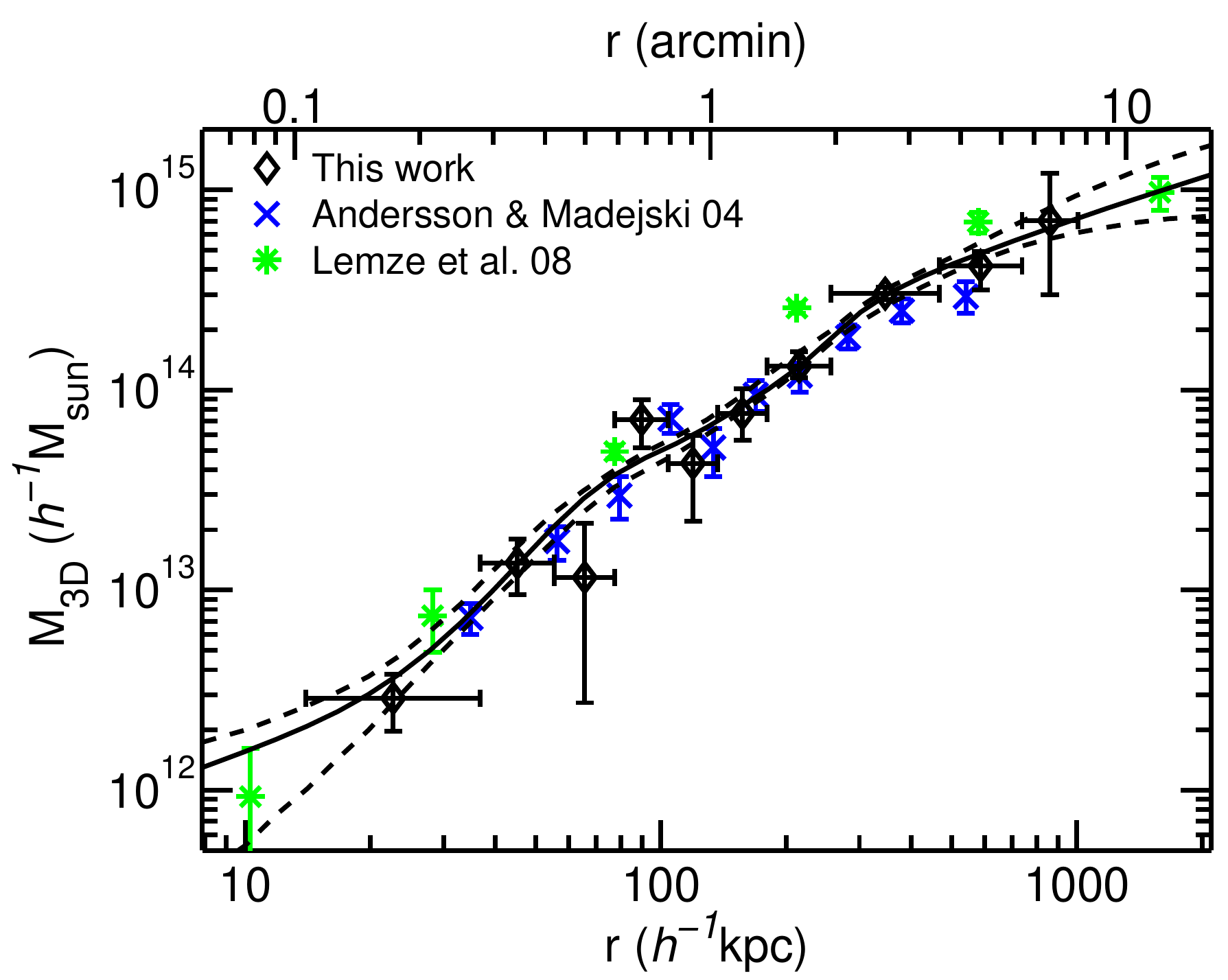}
\caption{The parametric mass profile (solid line) compared to the unparameterized result (diamonds). Dashed lines show the 95\% confidence bounds. Also shown are {\it XMM-Newton} result from \citetalias{and04} (crosses) and combined X-ray, strong and weak lensing analysis of \citetalias{lem07} (asterisks). The mass profile of \citetalias{lem07} is mainly determined by the lensing data.\label{plt_mtotpar}}
\end{center}
\end{figure}

Fig.\,\ref{plt_mtotpar} shows a comparison of parametric and non-parametric mass profiles. The non-parametric mass profile is from a more finely binned deprojected data than those shown in Fig.\,\ref{plt_proj}. Results from these two methods are fully consistent with each other, although their errors are quite different. For the non-parametric method, we simply assign the observable, e.g. $dT/dr$, at a certain radius, so the uncertainty associated with the position is not included in the error on the mass, $\sigma_M$, but separately shown on the radius. Therefore, $\sigma_M$ appears smaller if data are binned more coarsely. For the parametric method, the dependency of $\sigma_M$ on the data binning is weaker. The departure from the model for any data point is assumed to be random noise and is filtered out through the fitting. Hence, $\sigma_M$ reflects only the uncertainty of the fitted function and it depends strongly on the modeling.

\subsection{Comparison with other studies}\label{other}
\begin{figure}
\begin{center}
\includegraphics[width=8.5cm]{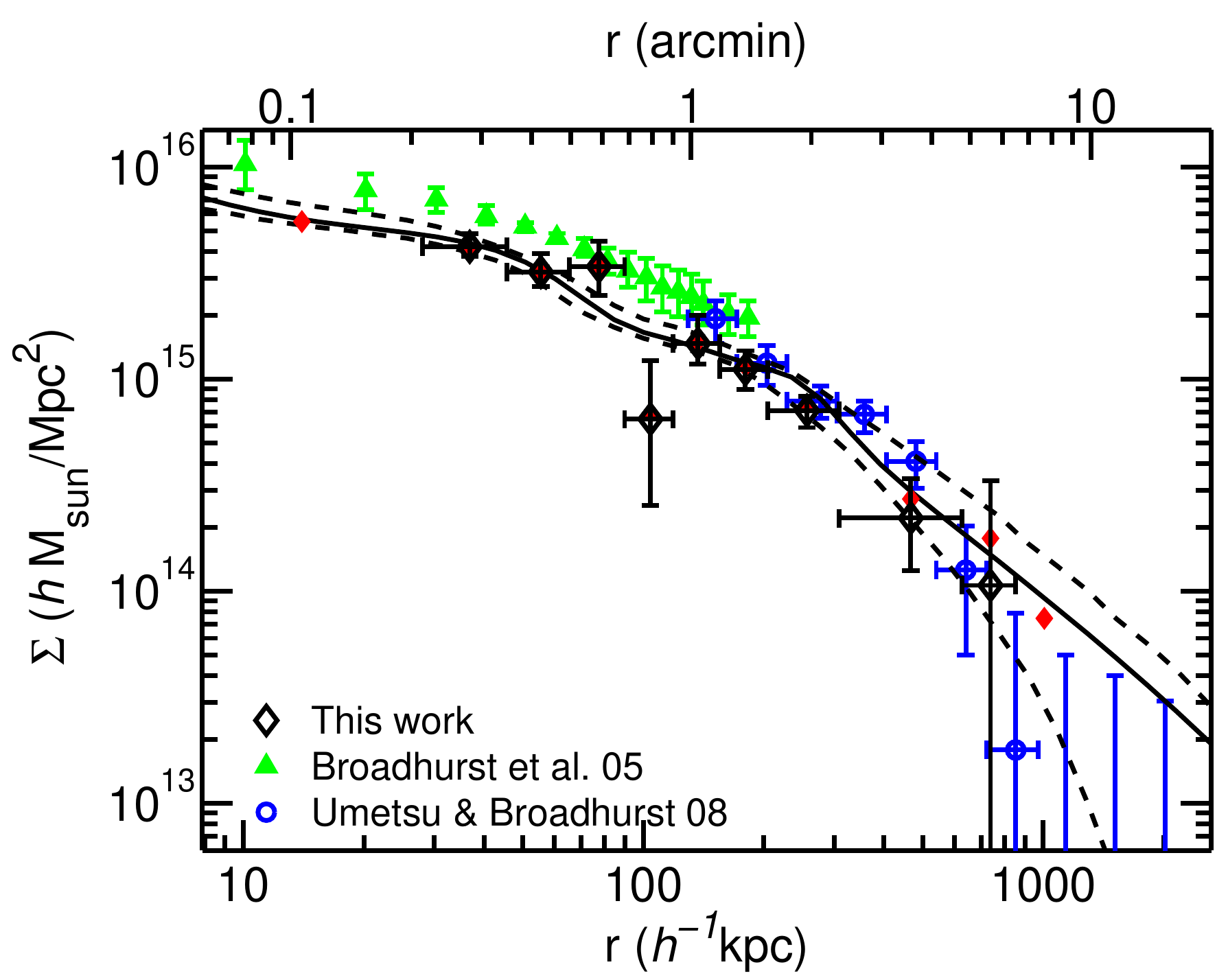}
\caption{Surface mass density profiles from non-parametric (open diamonds) and parametric X-ray model (solid and dashed lines, 95\% CL), compared to HST/ACS strong lensing analysis of \cite{bro05a} (triangles), and combined Subaru distortion and depletion data by \cite{ume08}, based on a maximum entropy method (circles). Filled diamonds show the mass from the nonparametric $T_{3D}$ and $n_e$ profiles that include estimations from the parametric result at 1\arcsec~(2 \kpc) and 13\arcmin~(1.7 \Mpc). 
\label{plt_sigma}}
\end{center}
\end{figure}

The total mass profiles of \citetalias{and04}, based on {\it XMM-Newton}, and \citetalias{lem07}, a joint X-ray, strong and weak lensing study are also shown in Fig.\,\ref{plt_mtotpar}. Our result is in good agreement with \citetalias{and04}, but disagrees with \citetalias{lem07} around $\sim$ 200 \kpc. To compare our mass estimate with other lensing works, we derived the total mass density and integrated it along the line-of-sight. The total mass density, $\rho$, is obtained through the hydrostatic equation, 
\begin{equation}\label{eq_massdensity}
4\pi G\rho=-\frac{k}{\mu_g m_p}(\nabla^2 T+T\nabla^2\mbox{ln}\rho_g+\nabla\mbox{ln}\rho_g\cdot\nabla T).
\end{equation}
For the nonparametric method, we evaluated Eq.~\ref{eq_massdensity} in a similar fashion as we did in \S \ref{nonparfit}. Errors were estimated from the Monte-Carlo simulations of deprojected $T$ and $\rho_g$ profiles. Fig.\,\ref{plt_sigma} shows the surface mass density profiles from both parametric and nonparametric methods, along with the HST/ACS strong lensing analysis of \cite{bro05a}, and the combined Subaru distortion and depletion data by \cite{ume08}. Since it requires at least 3 points to calculate the second derivative, $\rho$ at the boundary is unknown. This will introduce additional systematic errors to the inner and the outer projected profile. To demonstrate how this may affect our nonparametric result, we insert two artificial points at 1\arcsec~(2 \kpc) and 13\arcmin~(1.7 \Mpc) to the nonparametric $T$ and $\rho_g$ profiles with their values estimated from the parametric model. The projected density derived this way is shown in red filled diamonds in Fig.\,\ref{plt_sigma}. The X-ray data are consistent with those from the weak lensing, but disagree with the strong lensing analysis. Although the nonparametric data appears to agree with the strong lensing estimate at $r=80$ \kpc, this is probably due to the temperature fluctuation mentioned in \S \ref{parfit}. 

\begin{figure}
\begin{center}
\includegraphics[width=8.5cm]{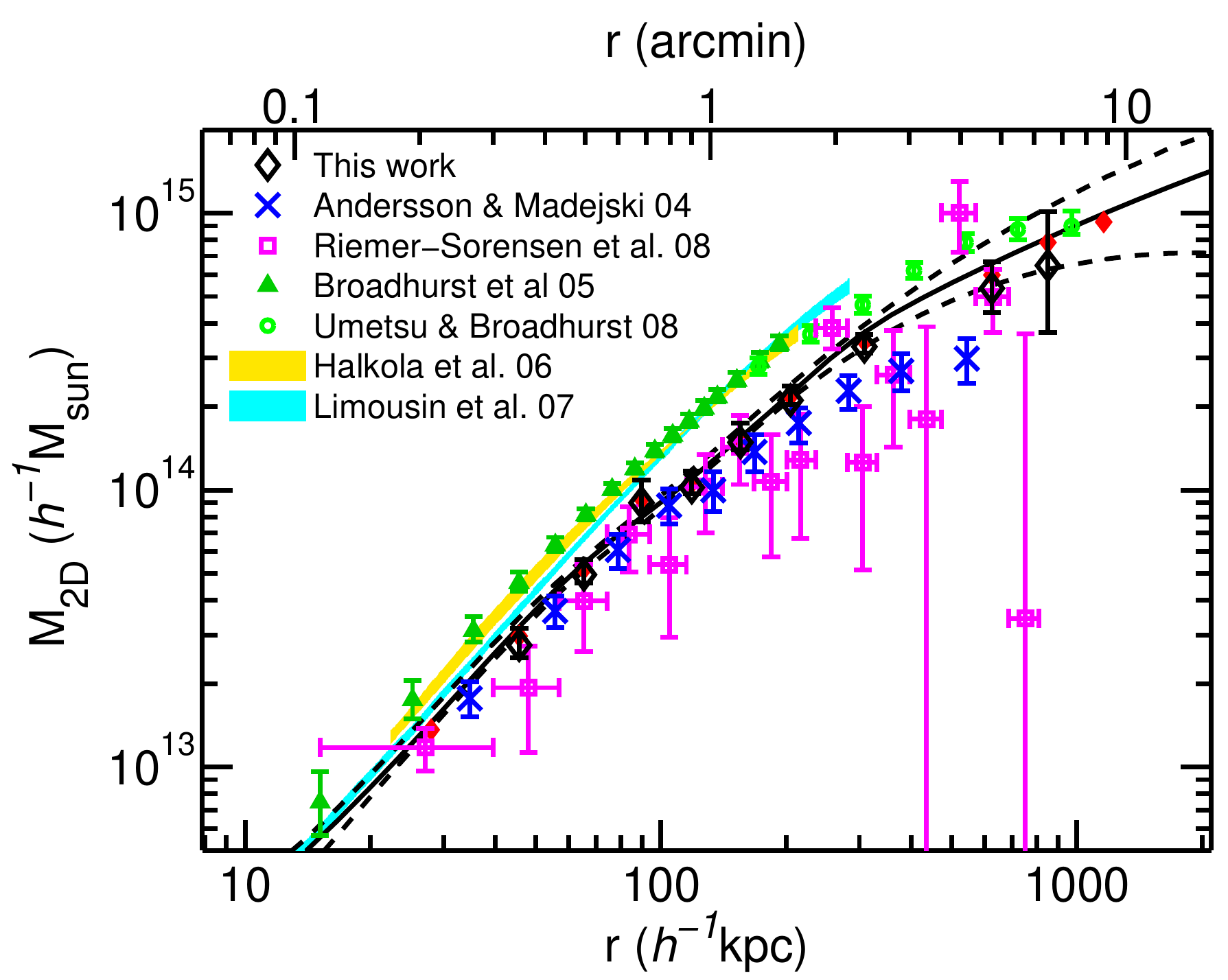}
\caption{Projected mass profiles from non-parametric (open diamonds) and parametric analyses (solid and dashed lines, 95\% CL), compared to {\it XMM-Newton} result from \citetalias{and04} (crosses), {\it Chandra} result by \cite{rie08} (squares), HST/ACS and Subaru results by \cite{bro05a} (triangles) and \cite{ume08} (circles). We integrated the lensing surface mass profile (shown in Fig.\,\ref{plt_sigma}) and estimated its uncertainties from Monte-Carlo simulations. Also shown are parametric strong lensing profiles of \cite{hal06} and \cite{lim07} (shaded regions, 68\% CL). \cite{rie08} used only SW part of the X-ray data and converted $M_{3D}$ to $M_{2D}$ with a NFW profile. \citetalias{and04} assumed that the last data point reached the cluster mass limit. Filled diamonds, same as Fig.\,\ref{plt_sigma}. \label{plt_m2d}}
\end{center}
\end{figure}

The mass discrepancy is manifested when comparing the cumulative projected mass profiles, $M_{2D}$, shown in Fig.\,\ref{plt_m2d}. The weak lensing $M_{2D}$ profile of \cite{ume08} includes the integration of the data of \cite{bro05a} in the inner region. Uncertainties are from Monte-Carlo simulations of the convergence profiles. The last 3 data points of \cite{ume08} (1-2.3 \Mpc) are discarded since only the upper limits are available. Also shown are parametric strong lensing profiles \citep{hal06,lim07}, and other X-ray analyses \citep[A04;][]{rie08}. To convert $M_{3D}$ to $M_{2D}$, \citetalias{and04} assume that the last data point reached the cluster mass limit, which unavoidably leads to underestimations especially at large radii.  \cite{rie08} use only the SE part of the cluster and four of the {\it Chandra} observations (excluding ObsID 540) and derive $M_{2D}$ based on a best-fit NFW model fit to the $M_{3D}$ profile. Their mass profile is generally lower than our estimate at most radii. This is contradictory to most findings that claim that the hydrostatic mass is underestimated in unrelaxed systems \citep[e.g.,][]{jel08}. Using such reasoning, and removing the NE part, presumably disturbed according to \cite{rie08}, should increase the overall mass estimate. The X-ray $M_{2D}$ is 25-40\% lower than that of lensing within 200 \kpc, corresponding to a $\sim 1.4\times10^{14}$ \hMsun~difference in the total projected mass. 

\subsection{NFW profile parameters}\label{nfw}
The total mass profile $M_{3D}$ was fit to the NFW model \citep{nfw97} to obtain the mass and the concentration parameter. To fit the nonparametric data, we weighted each point according to its vertical and horizontal errors, given by, 
\begin{equation}
\sigma^2=\sigma_M^2+\sigma_r^2\left(\frac{dM}{dr}\right)^2, 
\end{equation}
where $\sigma_r$ is assigned to be 68\% of the width of the horizontal error bar and $dM/dr$ is iteratively evaluated from the NFW model until it converges. In the parametric approach, a NFW model was fit to the parametrized mass profile that evaluated only at the radii where the projected temperature was measured with errors estimated from the standard deviation of a sample of mass profiles constructed from the simulated $T_{\text{proj}}$ and $S_X$ profiles described in \S \ref{parfit}. We repeated this procedure for all of the mass profiles in the sample. Resulting NFW parameters were used to estimate the uncertainty.  

Table\,\ref{tbl_nfw} lists the best-fit NFW parameters, $M_{200}$ and $c_{200}$, for the total mass from both methods and from other studies, all converted to the adopted cosmology. Compared to other X-ray studies, our derived $M_{200}$ is 30-50\% higher, closer to weak lensing results. The differences between our NFW parameters and those of \citetalias{and04} from {\it XMM-Newton} are primarily attributed to their slightly lower but yet consistent mass at the last data point (Fig.\,\ref{plt_mtotpar}). This demonstrates that the accurate mass measurement at large radii, where systematic errors are usually the greatest, is crucial to the determination of NFW parameters. 

Our results are consistent with weak lensing measurements, but with a lower concentration than what recent weak lensing studies seem to suggest \citep{ume08,cor09}. When these analyses are added with strong lensing information, a very tight constraint on the concentration parameter can be obtained, giving $C_{200}=9.9^{+ 0.8}_{-0.7}$ \citep{ume08}, which hardly can be reconciled with our value, $5.3^{+ 1.3}_{-1.2}$. However, if the gas emission is modeled with two spectral components with $\Th$ from \citetalias{lem07}, the X-ray derived concentration is in a closer agreement to those of combined strong and weak lensing studies, but this also implies that the majority of the gas is in the cool phase and occupies most of the intracluster space (\S\ref{2T}).

\begin{deluxetable*}{cccccl}
\tablewidth{0pt}
\tabletypesize{\scriptsize}
\tablecolumns{6} 
\tablecaption{Comparison of best-Fit NFW Parameters \label{tbl_nfw}}
\tablehead{\colhead{Method}     & 
           \colhead{Instrument}  &
           \colhead{$M_{200}$}    &  
           \colhead{$c_{200}$}  & 
           \colhead{$\chi^2/dof$}  &
           \colhead{Reference}\\
           \colhead{} & \colhead{} & \colhead{($10^{15}$\hMsun)} & \colhead{} & \colhead{} & \colhead{}}
\startdata
\sidehead{Spherical model}
X-ray (1T+edge)    & Chandra & $ 1.16^{+ 0.45}_{-0.27}$ &  $5.3^{+ 1.3}_{-1.2}$ &  6.3/8 & this work\\
X-ray (parametrized $T_{\text{3D}}$, $n_e$) & Chandra & $ 0.94^{+ 0.11}_{-0.06}$ & $ 6.6^{+ 0.4}_{-0.4}$ &  & this work\\ 
X-ray (2T\tna)         & Chandra & $ 1.45^{+ 0.36}_{-0.25}$ &  $7.6^{+ 1.3}_{-1.2}$ &  2.2/3 & this work\\
X-ray (2T\tna +edge)   & Chandra & $ 1.12^{+ 0.53}_{-0.29}$ &  $9.3^{+ 0.7}_{-2.8}$ &  0.1/3 & this work\\
X-ray (1T)       & XMM-Newton & $ 0.63\pm0.36$ &  $7.6^{+ 1.7}_{-2.6}$ &  7.6/8 & \citetalias{and04}\\
X-ray (1T)       & Chandra & 0.55  &  10.1\tnd &  1.6/13 & \cite{rie08}\\
SL               & ACS   & 2.29                       & $6.3^{+ 1.8}_{-1.6}$  &          & \cite{bro05a}\\
SL               & ACS   & $2.16\pm0.32$              & $5.8\pm0.5$           & 0.8/11   & \cite{hal06}\\ WL               & CFHT  & $0.97\pm0.13$              & $7.4\pm1.6$           &          & \cite{lim07}\\
WL               & CFHT  & $0.90\pm0.17$              & $13.1\pm7.5$   		  &          & \cite{cor09}\\   
WL                    & Subaru             & $1.24\pm0.14$ &  $10.5^{+ 4.4}_{-2.6}$ & 332/834 & \cite{ume08}\\
SL+WL                 & ACS+Subaru         & $1.22\pm0.13$ &  $10.8^{+ 1.1}_{-0.9}$ & 13.3/20 & \cite{bro05b}\\ 
SL+WL                 & ACS+Subaru         & $1.31\pm0.11$ &  $9.9^{+ 0.8}_{-0.7}$ & 335/846 & \cite{ume08}\\
SL+WL+X-ray ($S_X$)   & ACS+Subaru+Chandra & 1.42          &  $9.7^{+ 0.8}_{-0.7} $ & 15.3/24 & \citetalias{lem07}\\
\tableline
\sidehead{Triaxial model}
SL+WL\tnb             & ACS+Subaru & $1.15^{+ 0.26}_{-0.45}$  &  $13.4^{+ 1.8}_{-10.2}$          &  378/362 & \cite{ogu05}\\
WL\tnc                & CFHT  & $0.83\pm0.16$              & $12.0\pm6.6$  &          & \cite{cor09}\\ [-7pt]  
\enddata
\tablecomments{see \cite{com07,ume08,cor09} for a more complete compilation.}
\tablenotetext{a}{with $\Th$ from \citetalias{lem07}}
\tablenotetext{b}{under a flat prior on the axis ratios.}
\tablenotetext{c}{under a prior on the halo orientation that favors the line-of-sight direction.} 
\tablenotetext{d}{converted from best-fit parameters, $\rho_0=7.79\times10^6$\Msun kpc$^{-3}$, $r_s=174$ kpc ($h=0.7$, $\Omega _{m} = 0.28$, and $\Omega _{\lambda} = 0.72$), of \cite{rie08}, not consistent with their quoted value of 5.6 since they did not adopted the commonly defined $c_{200}$, the concentration at $r_{200}$ where the enclosed mean density is 200 times the critical density (private communication).}
\end{deluxetable*}

\subsection{Gas mass fraction}\label{fgas}
The cumulative gas fraction $f_{\text{gas}}=M_{\text{gas}}/M_{\text{total}}$, derived from our best-fit $T_{\text{3D}}$ and $n_e$ model, is $0.098^{+0.003}_{-0.004}$ $h_{70}^{-1.5}$ at $r_{2500}$ ($493^{+ 11}_{-10}$ \kpc), $\sim$20\% higher than what is found using  {\it XMM-Newton} data (\citetalias{and04}). In spite of this seemingly large difference, the data agree that $f_{\text{gas}}$ does not converge at $r_{2500}$. Much like in the case of A1689, the low-$z$ relaxed cluster A1413 does not have a strong cooling core and also has a steadily rising $f_{\text{gas}}$ profile out to $r_{500}$ \citep{pra02}. Comparing the $f_{\text{gas}}$ profile of A1413 with another nearby prominent cooling core cluster, A478, \cite{poi04} speculate that the flatter $f_{\text{gas}}$ profile of A478 is related to the presence of a cooling core. Our $f_{2500}$ is 11\% lower than the mean gas fraction of \cite{all08} derived from 42 relaxed clusters observed with {\it Chandra}, but our $f_{500}$, $0.12\pm0.01$ $h_{70}^{-1.5}$, agrees within 1\% of the $M-f_{\text{gas}}$ relation of \cite{vik08II}.

\section{DISCUSSION}\label{sec_dis}
\cite{nag07} show that following the data analysis of \cite{vik06}, the hydrostatic mass is underestimated by $14\pm6$\% within estimated $r_{2500}$ for simulated clusters visually classified as "relaxed". Based on the X-ray morphology, A1689 is likely to be categorized as a relaxed cluster. The X-ray centroid is within 3\arcsec~of the lensing and optical centers \citep{and04}, with a very minimal centroid shift or asymmetry \citep{has07}.  At the X-ray estimated $r_{2500}$ of 493 \kpc, we derive an enclosed hydrostatic mass of $(4.2\pm0.3)\times10^{14}$\hMsun, $\approx$ 30\% lower than the lensing mass from \citetalias{lem07}. At $r=200$ \kpc, this becomes a 50\% difference (see Fig.\,\ref{plt_mtotpar}). Such a strong bias is not seen in the relaxed cluster sample of \cite{nag07}, assuming that the lensing mass is unbiased, although this is not unusual for "unrelaxed" clusters, referring to those with secondary maxima, filamentary structures, or significant isophotal centroid shifts. 

\begin{deluxetable}{lcc}
\tablewidth{0pt}
\tabletypesize{\footnotesize}
\tablecaption{Comparison of $M_{500}$ \label{tbl_M500}}
\tablehead{\colhead{Method}     &
           \colhead{$M_{500}$}    &
           \colhead{$r_{500}$} \\
           \colhead{} & \colhead{($10^{14}$\hMsun)} & \colhead{(\Mpc)} }
\startdata
parametrized $T_{\text{3D}}$, $n_e$  & $7.3^{+1.3}_{-0.5}$  & $1.01^{+0.06}_{-0.03}$ \\
$M_{500}-T_X$\tndg                   & $7.7\pm0.2$          & $1.03\pm0.01$\\
$M_{500}-Y_X$\tndg\tna               & $7.7^{+0.5}_{-1.2}$  & $1.03^{+0.02}_{-0.06}$\\[-7pt]
\enddata
\tablenotetext{$\dagger$}{Scaling relations from \cite{vik08II} with indices fixed to self-similar theory values. Errors only reflect the measurement uncertainties. Dispersions of the relation is not included. $T_X=10.1\pm0.2$ keV, measured from $r=1.14\arcmin-7.6\arcmin$ ($\approx 0.15r_{500}-r_{500}$). }
\tablenotetext{a}{By solving Eq.~14 of \cite{vik08II}. The final $Y_X=T_X\times M_{\text{gas}}$ determined at $r_{500}$ is $(5.1^{+0.5}_{-1.2})\times10^{14}$ $h^{-2.5}$M$_\odot$ keV.}
\end{deluxetable}

Table\,\ref{tbl_M500} shows the comparison of measured $M_{500}$ with others derived from the $M_{500}-Y_X$ and $M_{500}-T_X$ relations of \cite{vik08II}, calibrated from 49 low-$z$ and 37 high-$z$ with $\langle z\rangle=0.5$ clusters observed with {\it Chandra} and {\it ROSAT}. A very good agreement has been achieved between these estimates. Since the $M_{500}-Y_X$ relation is insensitive to whether the cluster is relaxed or not \citep{kra06} and merging clusters tend to be cool for their mass \citep{mat01}, consistency among these mass estimates indicates that A1689 is relaxed in the sense that it behaves like other "relaxed" clusters on the scaling relation.

On the other hand, projection effects, such as triaxial halos or chance alignments, always have to be taken into account when comparing projected (lensing) and three-dimensional (X-ray) mass estimates. From kinematics of about 200 galaxies in A1689, \cite{lok06} suggest that there could be a few distant, possibly non-interacting, substructures superposed along the line of sight. \cite{lem08}, based on a $0.5\times0.5$ deg$^2$ VLT/VIMOS spectroscopic survey from \cite{czo04} which includes $\sim500$ cluster members, disagree with this projection view. They conclude that only one identifiable substructure at +3000 km/s, 1.5\arcmin~to the NE (The X-ray clump is at $\sim3.5$\arcmin~NE). This background group is seen in the strong lensing mass analysis \citep{bro05a}, but is determined not to be massive ($<10$\% of the total mass in the strong lensing region). Nonetheless, the higher than usual velocity dispersion in the cluster center, $\sim 2100$ km/s, indicates that the central part is quite complex \citep{czo04}. This may also imply that the halo is elongated in the line-of-sight direction, as galaxies move faster along the major axis.

For powerful strong lens systems, like A1689, halo sphericity is never a justified assumption \citep[e.g.,][]{hen07}. \cite{ogu08} show that these "superlens" clusters almost always have their major axes aligned along the line-of-sight, with more circular appearances in projection and $\sim40-60$\% larger concentrations than other clusters with similar masses and redshifts. \cite{gav05} demonstrates that using a prolate halo with axis ratio $\sim0.4$, they were able to explain the mass discrepancy between the lensing and X-ray estimates of cluster MS2137-23. This cluster has a well defined cool core \citep[e.g.,][]{and09}, thus presumably relaxed, and yet a factor of 2 difference in the mass is not lessened with a multiphase model for the core region \citep{ara04}. In contrast, triaxial modeling not only solves the mass inconsistency, but also the high concentration problem and the misalignment between stellar and dark matter components in MS2137-23 \citep{gav05}. 

\begin{figure}
\begin{center}
\includegraphics[width=8.5cm]{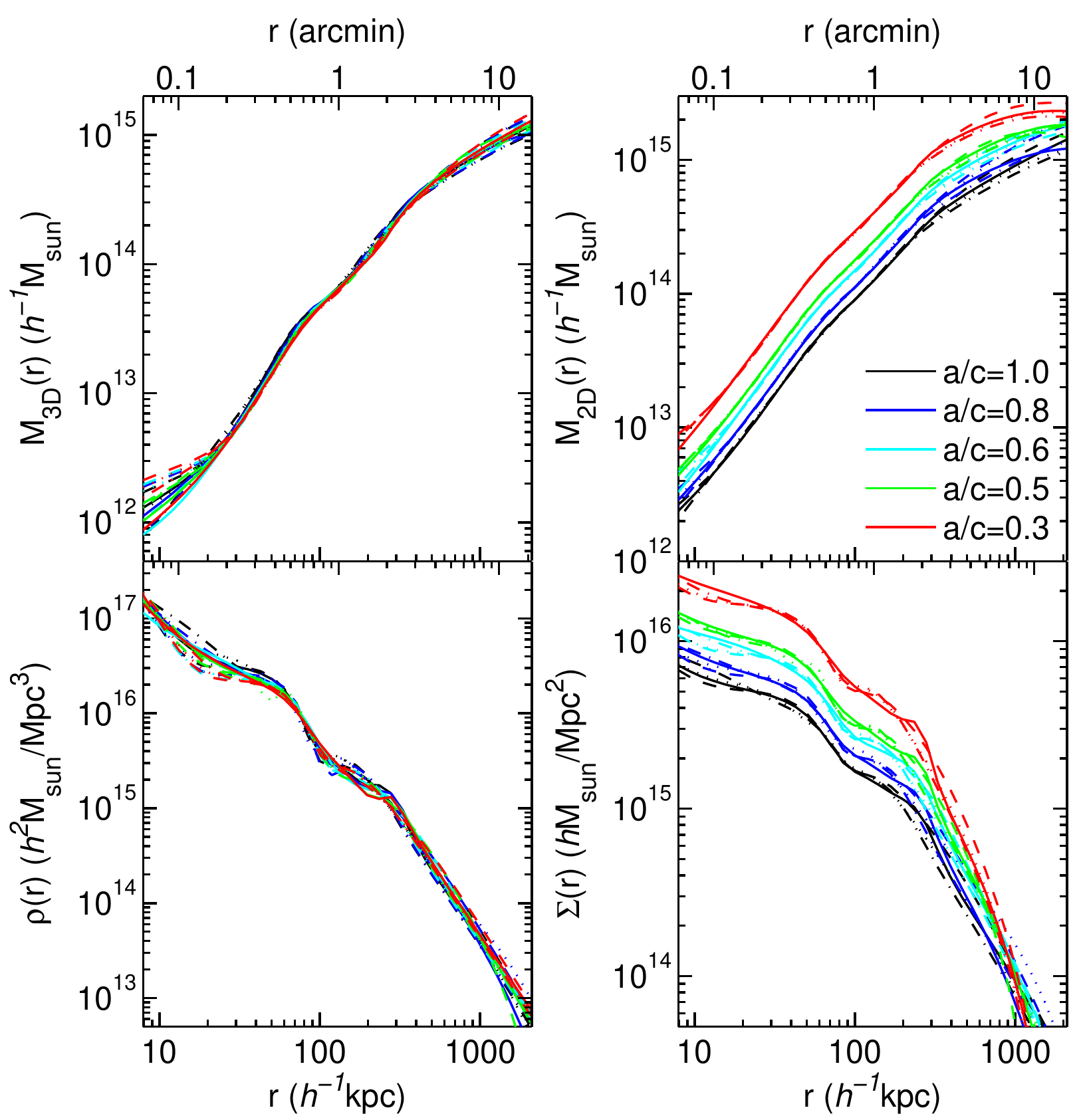}
\caption{Best-fit mass profiles for various axis ratios a/c from Model 1 (dash-dot line), 2 (dotted line), 3 (solid line), and 4 (dashed line). {\it Top left}: total mass enclose within a sphere of radius $r$, $M_{3D}(r)$. {\it Top right}: spherically average mass density $\rho(r)$. {\it Bottom left}: azimuthally average surface mass density $\Sigma(r)$. {\it Bottom right}: projected mass within a cylinder of radius $r$, $M_{2D}(r)$.
\label{plt_projmass}}
\end{center}
\end{figure}

To see how the triaxiality changes our mass measurements, we modeled $T_{\text{3D}}$ and $\rho_g$ with prolate profiles, by replacing $r$ in Eq.~\ref{eq_density} and \ref{eq_temp} with $(x^2/a^2+y^2/b^2+z^2/c^2)^{1/2}$, where we assumed $a=b<c$ and the major axis, $z$-axis, is perfectly aligned along the line-of-sight. Following the same analysis outlined in \S \ref{parfit} but with different projection factors, we obtained best-fit $T_{\text{3D}}$ and $\rho_g$ profiles. The derived mass profiles under various axis ratios $a/c$ are shown in Fig.\,\ref{plt_projmass}.  The uncertainties on $\Sigma(r)$ and $M_{2D}(r)$ are similar to those in Figs. \ref{plt_sigma} and \ref{plt_m2d}. We integrated the density from $z=-4.5$\Mpc~to +4.5\Mpc~($\approx 3 r_{200}$ for a/c=1) for all the cases. The uncertainties of $T_{\text{3D}}$ and $\rho_g$ profiles at large radii ($\gtrsim10$\% at $r=r_{500}$ and increasing further afterward) does not significantly change the projected mass at smaller radii ($\lesssim3$\% within 500\kpc).

The total mass enclosed within a sphere of radius $r$, $M_{3D}(r)$, and the spherically averaged mass density, $\rho(r)$, are basically unchanged under different assumptions of triaxiality, considering the typical measurement uncertainty. The same conclusion was drawn by \cite{pif03} and \cite{gav05}, though they assumed a $\beta$ or a NFW model with gas isothermality. For the azimuthally averaged surface mass density $\Sigma(r)$ or the projected mass within a cylinder of radius $r$, $M_{2D}(r)$, a factor of 2 or more difference can be easily made by increasing the ellipticity. An axis ratio of 0.6, giving $M_{2D}(<45\arcsec)=1.4\times10^{14}$\hMsun~(by a factor of 1.6 increase), can resolve the central mass discrepancy, but overpredicts the mass by $\sim$ 40\% at large radii. For a ratio of 0.7, the X-ray mass estimate data agrees with those of strong and weak lensing within 1\% ($-1\sigma$) and 25\% ($+1\sigma$), respectively. Since the gas distribution is rounder than that of the DM, a larger axis ratio than the finding of \cite{gav05} is expected.

Not only does the projected mass increases with the triaxiality, but also does the steepness of the profile. This explains a higher than X-ray derived concentration from the lensing data (\S \ref{nfw}). Although some attempts have been made to model the lensing mass profile with a 3D triaxial halo \citep{ogu05,cor09}, no significant constraint on the concentration parameter is obtained (Table\,\ref{tbl_nfw}). To break the degeneracy between the triaxiality and the concentration, observations from different prospective projections, such as X-ray, Sunyaev-Zel'dovich effect, or galaxy kinematics, are always needed.

\citet{rie08} conclude that A1689 harbors a cool core based on the radial temperature profile and a hardness ratio map. From this, they conclude that A1689 is relaxed, excluding the NE half of the cluster, where there is a low mass substructure.
Based on the derived temperature profile we disagree that A1689 contains a cool core. The temperature of the cluster varies radially from 9 to 11 keV with a slight drop only after 500 \kpc. This can not be characterized as the properties of a cool core cluster. In fact, as shown in \citet{and09}, A1689 is an intermediate stage cluster in terms of central baryon concentration with a minimal core temperature drop. This, does not necessarily provide evidence that the cluster is disturbed but we do not either expect the properties of a cool core cluster.
Hardness-ratio maps are very sensitive to accurate background subtraction, especially for high energy splittings. We suspect that the hardness ratio map ($S/H = E[0.3-6.0]/E[6.0-10.0]$) in \citet{rie08}, used as evidence for a cool core, suffers from residual background. The ratio decreases rapidly with radius from a central value of 2.2 and reaches 0.4 already at $3 \arcmin$. This is an extremely low ratio for any reasonable cluster temperature and it is in disagreement with the observed temperature profile. For comparison, a background-free spectrum from an isothermal cluster at $10~$keV would exhibit a count-ratio of $\sim 47$ in ACIS-I given the energy bands mentioned above. 
The usage of unsubtracted hardness-ratios in these bands shows that the high-energy band has a significant fractional background contribution and hence, is more spatially flat compared to the low energy band. This does not provide information about the spatial distribution of gas temperatures in the ICM. 
 
\section{SUMMARY}\label{sec_con}

We have investigated a deep exposure of Abell 1689 using the ACIS-I instrument aboard the Chandra X-ray telescope. In order to study the discrepancy of the gravitational mass from estimates from gravitational lensing, to that derived using X-ray data, we test the hypothesis of multiple temperature components in projection. 
The result of a two-temperature model fit shows that it is very important to take into account all details of the calibration of the instrument. We detect an additional absorption feature at 1.75 keV consistent with an absorption edge with an optical depth of 0.13. In analyzing multiple additional datasets, we find similar parameter values for this edge. 

If the edge is not modeled, fitting the cluster data within $3$\arcmin~strongly favors an additional plasma component at a different temperature. However, when this absorption feature is modeled, the second component does not improve the statistic significantly and the fit results is in better agreement with the XMM MOS data. In all cases, a second component has to have $T>5$ keV in order for the hot component to agree with the cluster temperature predicted by \citet{lem07} which is derived from lensing and $S_X$ profiles. This contradicts the assertion that cool clumps are biasing the X-ray temperature measurements since these substructures would not be cool at all. We also find that, if the temperature profile of the ambient cluster gas is in fact that of \citet{lem07}, the "cool clumps" would have to occupy 70-90\% of the space within $250~$kpc radius, assuming that the two temperature phases are in pressure equilibrium. In conclusion, we find the scenario proposed by \citet{lem07} unlikely.

Further studying the ratio of Fe \textsc{xxvi} K$\alpha$ and Fe \textsc{xxv} K$\alpha$ emission lines, we conclude that these show no signs of a multi-temperature projection and the best fit of this ratio implies a single temperature consistent with the continuum temperatures from both {\it XMM-Newton} MOS and the {\it Chandra} data when the absorption edge is modeled.

The discrepancy between lensing and X-ray mass estimates remains, particularly in the $r<200$ \kpc~region. Our X-ray mass profile shows consistent results compared to those from weak lensing \citep[e.g.,][]{bro05b,lim07,ume08,cor09}. Strong lensing mass profiles from different studies generally give consistent results \citep[e.g.,][]{bro05a,hal06,lim07}, but none of them agrees with those derived from X-ray observations \citep{xue02,and04,rie08}. Using a simple ellipsoidal modeling of the cluster with the major axis along the line of sight, we find that the projected mass, as derived from the X-ray analysis, increases by a factor of 1.6 assuming an axis-ratio of 0.6. We conclude that the mass discrepancy between lensing and X-ray derived masses can be alleviated by line of sight ellipticity and that this also can explain the high concentration parameter found in this cluster.

\acknowledgments
We thank the anonymous referee for valuable suggestions on the manuscript. Support
for this work was provided by NASA through SAO Award Number 2834-MIT-SAO-4018 issued by the 
Chandra X-Ray Observatory Center, which is operated by the Smithsonian Astrophysical Observatory 
for and on behalf of NASA under contract NAS8-03060. EP sincerely thanks John
Arabadjis for his support during early phases of this project.

\bibliographystyle{apj}

\begin{appendix}
\section{Instrumental absorption in {\it Chandra} data}
To see whether the absorption feature found in A1689 (\S\ref{system}) is a calibration problem, we analyzed other archived {\it Chandra} ACIS-I observations. Objects that have high-quality data and are relatively easy to model are pulsar wind nebula G021.5-00.9, elliptical galaxy NGC 4486 (M87), and the Coma cluster. Unfortunately, due to the high galactic absorption in G021.5-00.9, it is not suitable to use those observations to verify the instrumental absorption. Details of the datasets we used are listed in Table\,\ref{tbl_a_obslog}. These observations have low $N_H$ ($<3\times10^{20}$cm$^{-2}$), low background level ($<2$\% in 1.7-2.0 keV band), and high signal-to-noise ratios. Although the central part of M87 has very complex structures produced by the AGN \citep[e.g.,][]{for07}, the {\it XMM-Newton} observation indicates that the intracluster medium is likely to be single-phase in nature outside those regions \citep{mat02}. We extracted the spectra from $r=6\arcmin-7.5\arcmin$ of M87 and $r<5\arcmin$ of Coma cluster for each ACIS-I chip and fit with an absorbed single-temperature APEC model, multiplied by an absorption edge. The column density was fixed at the Galactic value \citep{dic90}. The redshift and all the elemental abundances, except Al, were free to vary. Parameters for this absorption edge are listed in Table\,\ref{tbl_a_edge}. For M87 whose emission is dominated by lines, these parameters are sensitive to the choice of the plasma model. Results from the latest MEKAL-based model, SPEX \citep[version 2.0;][]{kaa00}, are also shown in Table\,\ref{tbl_a_edge}. In general, an absorption at $\sim 1.75$ keV with an optical depth of 0.1-0.15 is seen in the datasets. However, for ObsID 7212, the absorption depth is determined less than 0.05 (1$\sigma$) on ACIS-I2 and I3. 

\begin{deluxetable*}{lcccccrccccccc}
\addtolength{\tabcolsep}{-2pt}
\tablewidth{0pt}
\tabletypesize{\footnotesize}
\tablecaption{{\it Chandra} Observation Log\label{tbl_a_obslog}}
\tablehead{
\colhead{Name}&
\colhead{$N_H$\tna}&
\colhead{z}&
\colhead{ObsID}&
\colhead{Data}&
\colhead{Obs. Date}&
\colhead{Exp.}&
\multicolumn{4}{c}{Background Norm.}&
\colhead{Region}&
\colhead{$f_B$\tnb}&
\colhead{$S/N$\tnc} \\ 
\colhead{     }&
\colhead{($10^{20}$cm$^{-2}$)}&
\colhead{     }&
\colhead{     }&
\colhead{Mode}&
\colhead{         }&
\colhead{(ks)    }&
\colhead{I0}&\colhead{I1}&\colhead{I2}&\colhead{I3}&
\colhead{     }&
\colhead{(\%)}&
\colhead{ }}
\startdata
M87 & 2.59 &0.00423 & \dataset [ADS/Sa.CXO\#obs/05826] {5826} & VFAINT & 2005-03-03 & 125.5 & 1.03 & 1.00 & 1.04 & 1.07 & 6\arcmin-7.5\arcmin & 1.5 & 235\\
    &      &        & \dataset [ADS/Sa.CXO\#obs/05827] {5827} & VFAINT & 2005-05-05 & 154.4 & 1.10 & 1.09 & 1.09 & 1.13 & & 1.5 & 262\\
    &      &        & \dataset [ADS/Sa.CXO\#obs/06186] {6186} & VFAINT & 2005-01-31 &  50.7 & 1.02 & 1.04 & 1.04 & 1.06 & & 1.5 & 150\\
    &      &        & \dataset [ADS/Sa.CXO\#obs/07212] {7212} & VFAINT & 2005-11-14 &  64.5 & 1.15 & 1.15 & 1.13 & 1.23 & & 1.7 & 168\\
Coma& 0.89 & 0.0231 & \dataset [ADS/Sa.CXO\#obs/09714] {9714} & VFAINT & 2008-03-20 & 29.6 & 1.32 & 1.34 & 1.18 & 1.43 & $<$5\arcmin& 1.9 & 134 \\[-7pt]
\enddata
\tablenotetext{a}{\cite{dic90}}
\tablenotetext{b}{Background fraction in 1.7-2.0 keV band.}
\tablenotetext{c}{Signal-to-noise ratio in 1.7-2.0 keV band.}
\end{deluxetable*}

\input{tab1.tex}
\end{appendix}
\end{document}

%% file: tab1.tex
\begin{table*}
\addtolength{\tabcolsep}{-2pt}
\footnotesize
\begin{center}
\caption{Absorption edge parameters\label{tbl_a_edge}}
\begin{tabular}{ccccccccc}
\hline\hline
     &       &     & \multicolumn{3}{c}{APEC}            & \multicolumn{3}{c}{SPEX} \\
Name & ObsID & ccd & $E_{thresh}$ & $\tau$  &z           & $E_{thresh}$ & $\tau$  &z          \\
      &     &      & (keV)        &         &            & (keV)        &         &           \\
\hline
M87 &5826 & 0 & $1.47^ { +0.02 } _ { -0.02 } $ & $0.08^ { +0.02 } _ { -0.02 } $ & $(4.8^ { +0.5 } _ { -0.6 } )\times10^{-3}$ & $1.47^ { +0.02 } _ { -0.02 } $ & $0.10^ { +0.02 } _ { -0.02 } $ & $(4.0^ { +0.7 } _ { -0.7 } )\times10^{-3}$  \\
 &     & 1 & $1.73^ { +0.02 } _ { -0.03 } $ & $0.07^ { +0.02 } _ { -0.02 } $ & $(5.1^ { +0.5 } _ { -0.5 } )\times10^{-3}$ & $1.70^ { +0.02 } _ { -0.03 } $ & $0.08^ { +0.01 } _ { -0.02 } $ & $(4.4^ { +0.4 } _ { -0.4 } )\times10^{-3}$  \\
 &     & 2 & $1.75^ { +0.01 } _ { -0.02 } $ & $0.12^ { +0.03 } _ { -0.02 } $ & $(12.8^ { +0.6 } _ { -1.0 } )\times10^{-3}$ & $1.73^ { +0.01 } _ { -0.02 } $ & $0.12^ { +0.02 } _ { -0.02 } $ & $(8.4^ { +1.2 } _ { -0.3 } )\times10^{-3}$  \\
 &     & 3 & $1.79^ { +0.02 } _ { -0.02 } $ & $0.09^ { +0.02 } _ { -0.02 } $ & $(5.3^ { +0.2 } _ { -0.3 } )\times10^{-3}$ & $1.75^ { +0.03 } _ { -0.02 } $ & $0.08^ { +0.02 } _ { -0.02 } $ & $(4.5^ { +0.2 } _ { -0.3 } )\times10^{-3}$ \\
 &5827 & 0 & $1.75^ { +0.02 } _ { -0.02 } $ & $0.11^ { +0.03 } _ { -0.02 } $ & $(4.9^ { +0.8 } _ { -0.7 } )\times10^{-3}$ & $1.72^ { +0.02 } _ { -0.01 } $ & $0.11^ { +0.01 } _ { -0.02 } $ & $(3.8^ { +1.0 } _ { -0.8 } )\times10^{-3}$  \\
 &     & 1 & $1.73^ { +0.02 } _ { -0.02 } $ & $0.10^ { +0.02 } _ { -0.02 } $ & $(5.5^ { +0.5 } _ { -1.3 } )\times10^{-3}$ & $1.72^ { +0.02 } _ { -0.02 } $ & $0.09^ { +0.01 } _ { -0.02 } $ & $(3.1^ { +0.3 } _ { -0.7 } )\times10^{-3}$  \\
 &     & 2 & $1.81^ { +0.02 } _ { -0.02 } $ & $0.11^ { +0.03 } _ { -0.02 } $ & $(4.1^ { +0.5 } _ { -0.2 } )\times10^{-3}$ & $1.74^ { +0.03 } _ { -0.04 } $ & $0.08^ { +0.02 } _ { -0.02 } $ & $(2.4^ { +0.3 } _ { -0.4 } )\times10^{-3}$  \\
 &     & 3 & $2.14^ { +0.03 } _ { -0.04 } $ & $0.06^ { +0.02 } _ { -0.02 } $ & $(3.8^ { +0.5 } _ { -0.3 } )\times10^{-3}$ & $2.13^ { +0.04 } _ { -0.06 } $ & $0.06^ { +0.02 } _ { -0.02 } $ & $(2.3^ { +0.5 } _ { -0.3 } )\times10^{-3}$  \\
 &6186 & 0 & $1.72^ { +0.06 } _ { -0.06 } $ & $0.09^ { +0.03 } _ { -0.05 } $ & $(4.6^ { +0.7 } _ { -0.6 } )\times10^{-3}$ & $1.70^ { +0.04 } _ { -0.04 } $ & $0.09^ { +0.02 } _ { -0.03 } $ & $(4.0^ { +0.7 } _ { -0.7 } )\times10^{-3}$  \\
 &     & 1 & $1.73^ { +0.03 } _ { -0.04 } $ & $0.11^ { +0.04 } _ { -0.03 } $ & $(5.1^ { +0.7 } _ { -0.5 } )\times10^{-3}$ & $1.71^ { +0.03 } _ { -1.71 } $ & $0.11^ { +0.03 } _ { -0.03 } $ & $(4.4^ { +0.8 } _ { -0.6 } )\times10^{-3}$  \\
 &     & 2 & $1.72^ { +0.07 } _ { -0.06 } $ & $0.11^ { +0.05 } _ { -0.03 } $ & $(8.2^ { +0.5 } _ { -0.4 } )\times10^{-3}$ & $1.66^ { +0.03 } _ { -0.10 } $ & $0.10^ { +0.02 } _ { -0.02 } $ & $(7.9^ { +0.4 } _ { -0.7 } )\times10^{-3}$  \\
 &     & 3 & $1.74^ { +0.03 } _ { -0.03 } $ & $0.12^ { +0.03 } _ { -0.03 } $ & $(8.1^ { +0.3 } _ { -0.4 } )\times10^{-3}$ & $1.70^ { +0.04 } _ { -0.03 } $ & $0.11^ { +0.03 } _ { -0.03 } $ & $(7.8^ { +0.3 } _ { -0.3 } )\times10^{-3}$  \\
 &7212 & 0 & $1.77^ { +0.03 } _ { -0.02 } $ & $0.11^ { +0.02 } _ { -0.04 } $ & $(4.3^ { +0.8 } _ { -0.6 } )\times10^{-3}$ & $1.62^ { +0.09 } _ { -1.62 } $ & $0.08^ { +0.02 } _ { -0.03 } $ & $(1.0^ { +3.4 } _ { -1.0 } )\times10^{-3}$  \\
 &     & 1 & $1.91^ { +0.21 } _ { -0.05 } $ & $0.10^ { +0.04 } _ { -0.04 } $ & $(5.8^ { +0.4 } _ { -0.3 } )\times10^{-3}$ & $2.00^ { +0.04 } _ { -0.04 } $ & $0.10^ { +0.03 } _ { -0.05 } $ & $(2.0^ { +0.4 } _ { -0.4 } )\times10^{-3}$  \\
 &     & 2 & $1.74 _ { f }  $ & $0.01^ { +0.03 } _ { -0.01 } $ & $(6.4^ { +0.5 } _ { -0.7 } )\times10^{-3}$ & $1.74 _ { f }  $ & $0.02^ { +0.03 } _ { -0.02 } $ & $(1.2^ { +1.0 } _ { -1.0 } )\times10^{-3}$  \\
\hline
Coma &9714 & 0 & $1.76^ { +0.03 } _ { -0.02 } $ & $0.14^ { +0.06 } _ { -0.02 } $ & $(4.6^ { +0.9 } _ { -0.2 } )\times10^{-2}$ &
 $1.76^ { +0.03 } _ { -0.02 } $ & $0.15^ { +0.04 } _ { -0.05 } $ & $(4.6^ { +0.8 } _ { -0.2 } )\times10^{-2}$ \\
 &     & 1 & $1.75^ { +0.05 } _ { -0.05 } $ & $0.08^ { +0.04 } _ { -0.05 } $ & $(3.1^ { +0.5 } _ { -0.6 } )\times10^{-2}$ & $1.75^ { +0.05 } _ { -0.05 } $ & $0.08^ { +0.03 } _ { -0.05 } $ & $(2.6^ { +0.8 } _ { -0.4 } )\times10^{-2}$  \\
 &     & 2 & $1.75^ { +0.02 } _ { -0.02 } $ & $0.11^ { +0.04 } _ { -0.02 } $ & $(3.2^ { +0.7 } _ { -0.3 } )\times10^{-2}$ & $1.75^ { +0.03 } _ { -0.03 } $ & $0.10^ { +0.05 } _ { -0.02 } $ & $(3.2^ { +0.6 } _ { -0.4 } )\times10^{-2}$  \\
 &     & 3 & $1.77^ { +0.02 } _ { -0.02 } $ & $0.18^ { +0.03 } _ { -0.03 } $ & $(3.8^ { +0.2 } _ { -0.4 } )\times10^{-2}$ & $1.77^ { +0.02 } _ { -0.02 } $ & $0.17^ { +0.03 } _ { -0.03 } $ & $(3.7^ { +0.3 } _ { -0.5 } )\times10^{-2}$  \\
 &     & 3 & $1.75^ { +0.06 } _ { -0.06 } $ & $0.03^ { +0.03 } _ { -0.02 } $ & $(4.8^ { +0.5 } _ { -0.5 } )\times10^{-3}$ & $1.73^ { +0.05 } _ { -0.09 } $ & $0.03^ { +0.02 } _ { -0.01 } $ & $(2.3^ { +0.3 } _ { -0.8 } )\times10^{-3}$ \\
\hline
A1689 &540 & 13 & $1.72^ { +0.05 } _ { -0.03 } $ & $0.10^ { +0.03 } _ { -0.04 } $ & $0.190^ { +0.002 } _ { -0.002 } $ & $1.72^ { +0.05 } _ { -0.04 } $ & $0.09^ { +0.03 } _ { -0.03 } $ & $0.189^ { +0.004 } _ { -0.003 } $ \\
&1663 & 3 & $1.75^ { +0.04 } _ { -0.03 } $ & $0.16^ { +0.04 } _ { -0.03 } $ & $0.191^ { +0.006 } _ { -0.011 } $ & $1.75^ { +0.04 } _ { -0.03 } $ & $0.16^ { +0.04 } _ { -0.04 } $ & $0.192^ { +0.005 } _ { -0.012 } $  \\
&5004 & 3 & $1.76^ { +0.04 } _ { -0.03 } $ & $0.15^ { +0.02 } _ { -0.03 } $ & $0.182^ { +0.002 } _ { -0.005 } $ & $1.76^ { +0.03 } _ { -0.03 } $ & $0.15^ { +0.03 } _ { -0.03 } $ & $0.180^ { +0.003 } _ { -0.003 } $  \\
&6930 & 3 & $1.74^ { +0.01 } _ { -0.02 } $ & $0.12^ { +0.01 } _ { -0.01 } $ & $0.188^ { +0.003 } _ { -0.003 } $ & $1.75^ { +0.01 } _ { -0.01 } $ & $0.12^ { +0.01 } _ { -0.01 } $ & $0.188^ { +0.002 } _ { -0.002 } $  \\
&7289 & 3 & $1.73^ { +0.02 } _ { -0.02 } $ & $0.14^ { +0.01 } _ { -0.02 } $ & $0.183^ { +0.002 } _ { -0.003 } $ & $1.73^ { +0.02 } _ { -0.02 } $ & $0.13^ { +0.01 } _ { -0.01 } $ & $0.181^ { +0.003 } _ { -0.003 } $  \\
\hline
\end{tabular}
\end{center}
\end{table*}